\documentclass[prc,aps,twocolumn]{revtex4}
\usepackage{graphicx}
\usepackage{dcolumn}
\usepackage{bm}

\def\tend{\mathop{\to}}

\def\lim{\mathop{\rm {lim}}}

\begin{document}
\widetext
\title{Generalized dynamical equation and low energy nucleon dynamics}
\author{Renat Kh.Gainutdinov and Aigul A.Mutygullina}
\affiliation{Department of Physics, Kazan State University, 18
Kremlevskaya St, Kazan 420008, Russia}
\date{\today}
 
\begin{abstract}
Low energy nucleon dynamics is investigated by using the
generalized dynamical equation derived in [J. Phys. A v.32, 5657
(1999)]. This equation extends quantum dynamics to describe the
time evolution in the case of  nonlocal-in-time interactions. We
show that the use of the generalized dynamical equation, which
allows one to take into account that effective actions in quantum
field theory, arising after integrating out certain degrees of
freedom, are generally nonlocal, provides a new way to formulate
the effective theory of nuclear forces.  In this way we construct
the most general possible operator describing the short-range part
of the $NN$ interaction in the ${}^1S_0$ channel. In particular,
this operator can be incorporated into the chiral potential model.
This is shown to provide an extension of the standard chiral
approach. Being equivalent to the chiral potential model for some
values of the constants characterizing the short-range of the $NN$
interaction this approach allows the solutions that cannot be
reproduced by any potential model. In this way we find that for
certain values of the constants inconsistent with potentials
models the effect of the short-range forces on the half off-shell
$2N$ $T$-matrix with high off-shell momenta grows in magnitude
rapidly as the kinetic energy  of outgoing protons decreases at
laboratory energies below $15$ $MeV$.  We show that this effect
may be the origin of the existing discrepancy between the theory
and experiment in describing the proton-proton bremsstrahlung.
\end{abstract}
 \maketitle
\narrowtext

\section{Introduction}
Ideas from the foundations of quantum mechanics are being applied
now to many branches of physics. In the quantum mechanics of
particles interacting through the Coulomb potential one deals with
a well-defined interaction Hamiltonian and the Schr{\"o}dinger
equation governing the dynamics of the theory. This theory is
perfectly consistent and provides an excellent description of
atomic phenomena at low energies. It is natural to expect that low
energy nuclear physics can be described in the same way. However,
one has not yet constructed a fundamental nucleon-nucleon (NN)
potential. Nowadays there exist phenomenological $NN$ potentials
which successfully describe scattering data to high precision, but
they do not emerge from QCD and contain $ad$ $hoc$ form factors. A
first attempt to systematically solve the problem of low energy
nucleon dynamics and construct a bridge to QCD was made by
Weinberg \cite{EFT3}. He suggested to derive a $NN$ potential in
time-ordered chiral perturbation theory (ChPT). However, such a
potential is singular and the Schr{\"o}dinger (Lippmann-Schwinger)
equation does not make sense without regularization and
renormalization. This means that in the effective field theory
(EFT) of nuclear forces, which following the pioneering work of
Weinberg has become very popular in nuclear physics (for a review,
see Ref. \cite{rev}), the Schr{\"o}dinger equation is not valid.
On the other hand, the whole formalism of fields and particles can
be considered as an inevitable consequence of quantum mechanics,
Lorentz invariance, and the cluster decomposition principle
\cite{Weinberg97}. Thus in the nonrelativistic limit QCD must
reproduce low energy nucleon physics consistent with the basic
principles of quantum mechanics. However, as it follows from the
Weinberg analysis, QCD leads through ChPT to the low energy theory
in which the Schr{\"o}dinger equation is not valid. This means
that either there is something wrong with QCD and ChPT or the
Schr{\"o}dinger equation is not the basic dynamical equation of
quantum theory. Meanwhile, in Ref. \cite{R.Kh.:1999} it has been
shown that the Schr{\"o}dinger equation is not the most general
equation consistent with the current concepts of quantum physics,
and a more general equation of motion has been derived as a
consequence of the basic postulates of the Feynman
\cite{Feynman:1948} and canonical approaches to quantum theory.
Being equivalent to the Schr{\"o}dinger equation in the case of
instantaneous interactions, this generalized dynamical equation
permits the generalization to the case where the dynamics of a
system is generated by a nonlocal-in-time interaction. The
generalized quantum dynamics (GQD) developed in this way has
proved an useful tool for solving various problems in quantum
theory \cite{PRC:2002,PLA:2002}.

The  GQD allows one to consider the problem of consistency of
quantum mechanics with the low energy predictions of QCD from a
new  point of view. From this viewpoint, in investigating the
consequences of ChPT we must not restrict ourselves to the
assumption that the $NN$ interaction can be parametrized by a $NN$
potential, and low energy nucleon dynamics is governed by the
Schr{\"o}dinger equation. This dynamics may be governed by the
generalized dynamical equation (GDE) with a nonlocal-in-time
interaction operator when this equation is not equivalent to the
Schr{\"o}dinger equation, and hence the above divergence problems
may be the cost of trying to describe low energy nucleon dynamics
in terms of Hamiltonian formalism while this dynamics is really
non-Hamiltonian. Another motivation for investigating the problem
of the description of nucleon dynamics from the point of view of
the formalism of the GQD is still existing discrepancy between the
predictions of the modern $NN$ potential models and experiment.
For example, there is a significant discrepancy between theory and
experiment in describing the proton-proton ($pp$) bremsstrahlung
\cite{M-S} that is an accurate test of $NN$ interaction models. An
important part of this discrepancy originates in a poor
description of the $NN$ interaction at low energies \cite{C-T}.
The formalism of the GQD that reproduces the potential models of
the $NN$ interaction in the particular case when this interaction
is assumed to be instantaneous admits a more general class of such
models, and only experiment such as the $pp$ bremsstrahlung can
discriminate between them. In this context the discrepancy between
the predictions of the potential models and experiment should mean
that in describing nucleon dynamics one cannot ignore that the
$NN$ interaction is nonlocal in time.

The aim of the present paper is twofold. From the one hand, we
intend to show that the low energy predictions of QCD are
consistent with the basic principles of quantum mechanics, but in
this case a new insight into these principles provided by the
formalism of the GQD is needed. On the other hand, it is our
intention to demonstrate that this formalism allows one to
formulate the effective theory of nuclear forces as an inevitable
consequence of the basic principles of quantum mechanics and the
symmetries of QCD. Being formulated in this way the theory of
nuclear forces opens new possibilities for describing nucleon
dynamics at low energies. In Sec. II we briefly consider the main
features of the formalism of the GQD developed by one of the
authors (R.G.) in Ref. \cite{R.Kh.:1999}. We mainly focus on the
physical meaning of the GDE which is a direct consequence of the
principle of the superposition of the probability amplitudes and
the requirement that the evolution operator is unitary. This
equation plays a key role in the present work.

In Sec. III we consider the pionless theory of nuclear forces. By
focusing the attention on the ${}^1S_0$  channel, we show that the
requirement that the two-nucleon $(2N)$ $T$-matrix satisfies the
GDE and is consistent with the symmetries of QCD allows one to
construct this $T$-matrix by expanding it in powers of $Q/\Lambda$
where $Q$ being some low energy scale, and $\Lambda$ is the scale
at which the theory is expected to break down.  In this way we can
reproduce all results for the $NN$ scattering amplitudes obtained
in the standard EFT of nuclear forces. It is important that these
results are reproduced starting with a well defined $2N$
$T$-matrix without resorting to the regularization and
renormalization procedures. In Sec. IV we show that the use of the
GDE allows one to formulate the effective theory as a perfectly
consistent theory free from UV divergences. Being formulated in
this way, the effective theory keeps all
 advantages
of the traditional nuclear physics approach. At the same time, its
advantage over the traditional approach is that it allows one to
find constraints on the off-shell behavior of the $2N$ $T$-matrix
that are placed by the symmetries of QCD.

In Sec. V the proposed formalism is considered from the point of
view of the Weinberg program for physics of the two-nucleon
systems. The approach based on the derivation of a $NN$ potential
from ChPT was pioneered by Weinberg \cite{Weinberg97} and
developed by Ord{\^o}nez \cite{Ordonez} and van Kolck \cite{van
Kolck}. Recently the potential at fourth order of ChPT that
reproduce the $NN$ data with the same accuracy as phenomenological
high-precision potentials was developed by Entem and Machleidt
\cite{Entem2}. We show that our formulation of the effective
theory of nuclear forces is a generalization of the above
approach. Instead of the Schr{\"o}dinger (LS) equation, it is
suggested to use the GDE. In this way we construct the most
general possible operator describing the short-range part of the
$NN$ interaction. We show that this operator can be incorporated
into the chiral approach. Being equivalent to the chiral potential
model for some values of the constants characterizing the
short-range part of the $NN$ interaction this approach allows the
solutions that cannot be reproduced by any potential model. In
Sec. VI we show that the for certain values of the constants
inconsistent with potentials models the effect of the short-range
forces on the $2N$ $T$-matrix with high off-shell momenta grows in
magnitude rapidly as the kinetic energy  of outgoing protons
decreases in the region below $15$ $MeV$. We also show that the
fact that the existing microscopic $pp$ bremsstrahlung models
cannot reproduce this effect may be the origin of the discrepancy
between their predictions and the experimental data. We also show
that our formalism allows one to modify the short-range parts of
the existing high-precision potentials to improve their off-shell
predictions keeping the phase shifts unchanged.  New possibilities
that the formalism of the GQD opens for solving three-body problem
in nuclear physics are discussed in Sec. VII.

\section{Generalized quantum dynamics}

The basic concept of the canonical formalism of quantum theory is
that it can be formulated in terms of vectors of a Hilbert space
and operators acting on this space. In this formalism the
postulates that establish the connection between the vectors and
operators and states of a quantum system and observables are used
together with the dynamical postulate according to which the time
evolution of a quantum system is governed by the Schr{\"o}dinger
equation. In the Feynman formalism quantum theory is formulated in
terms of probability amplitudes without resorting to vectors and
operators acting on a Hilbert space. Feynman's theory starts with
an analysis of the phenomenon of quantum interference. The results
of this analysis which leads directly to the concept of
 the superposition of probability amplitudes are summarized by
the following postulate \cite{Feynman:1948}:

\textit{ The probability of an event is the absolute square of a
complex number called the probability amplitude. The joint
probability amplitude of a time-ordered sequence of events is the
product of separate probability amplitudes of each of these
events. The probability amplitude of an event which can happen in
several different ways is a sum of the probability amplitudes for
each of these ways.}

The Feynman formulation is based on the assumption that the
history of a quantum system can be represented by some path in
space-time, and hence the probability amplitude of any event is a
sum of the probability amplitudes that a particle has a completely
specified path in space-time. The contribution from a single path
is postulated to be an exponential whose (imaginary) phase is the
classical action (in units of $\hbar$) for the path in question.
This assumption is not as fundamental as the above principle of
the superposition of probability amplitudes which follows directly
from the analyzes of the phenomenon of quantum interference. This
fact is emphasized, for example, in Feynman's book \cite{Feyn},
where a minimal set of physical principles which must be satisfied
in any theory of fields and particles is analyzed. Feynman
includes in this set neither the second postulate of his formalism
nor the Schr{\"o}dinger equation: The only  quantum mechanical
principle included in this set is the principle of the
superposition of probability amplitudes. In Ref. \cite{R.Kh.:1999}
it has been shown that, instead of processes associated with a
completely specified path in space-time, one can use processes
associated with completely specified instants of the beginning and
end of interaction in a quantum system as alternative ways in
which any event can happen. As it turned out, employing this class
of alternatives allows one to derive a dynamical equation from the
principle of the superposition without making supplementary
assumptions like the second postulate of Feynman's theory.
 By using this class of
alternatives and the superposition principle,
$\langle\psi|U(t,t_0)|\varphi\rangle$, being the probability
amplitude of finding a quantum system in the state $|\psi\rangle$
in a measurement at time $t$,
 if at time $t_0$ it was in the state
 $|\varphi\rangle$, can  be represented in the form \cite{R.Kh.:1999}
\begin{eqnarray}
\langle\psi| U(t,t_0)|\varphi\rangle =
\langle\psi|\varphi\rangle 
+ \int_{t_0}^t dt_2 \int_{t_0}^{t_2} dt_1 \langle\psi|\tilde
S(t_2,t_1)|\varphi\rangle. \label{ev}
\end{eqnarray}
Here $\langle\psi|\tilde S(t_2,t_1)|\varphi\rangle$ is the
probability amplitude that, if at time $t_1$ the system was in the
state $|\varphi\rangle,$ then the interaction in the system will
begin at time $t_1$ and  end at time $t_2$ and at this time the
system will be in the state $|\psi\rangle.$ By using the operator
formalism, one can represent amplitudes $\langle\psi|
U(t,t_0)|\varphi\rangle$ by the matrix elements of the unitary
evolution operator $U(t,t_0)$ in the interaction picture. The
operator $\tilde S(t_2,t_1)$ represents the contribution to the
evolution operator from the process in which the interaction in
the system begins at time $t_1$ and ends at time $t_2$.  As has
been shown in Ref. \cite{R.Kh.:1999}, for the evolution operator
(\ref{ev}) to be unitary for any $t$ and $t_0$ the operator
$\tilde S(t_2,t_1)$ must satisfy the equation
\begin{eqnarray}
&&(t_2-t_1) \tilde S(t_2,t_1) \nonumber\\
&=& \int^{t_2}_{t_1} dt_4 \int^{t_4}_{t_1}dt_3 (t_4-t_3) \tilde
S(t_2,t_4) \tilde S(t_3,t_1). \label{main}
\end{eqnarray}
A remarkable feature of this equation is that it works as a
recursion relation and allows one to obtain the operators $\tilde
S(t_2,t_1)$ for any $t_1$ and $t_2$, if $\tilde S(t'_2, t'_1)$
corresponding to infinitesimal duration times $\tau = t'_2 -t'_1$
of interaction are known. It is natural to assume that most of the
contribution to the evolution operator in the limit $t_2\to t_1$
comes from the processes associated with the fundamental
interaction in the system under study. Denoting this contribution
by $H_{int}(t_2,t_1)$ we can write
\begin{equation}
\tilde{S}(t_2,t_1) \tend\limits_{t_2\rightarrow t_1}
H_{int}(t_2,t_1) + o(\tau^{\epsilon}),\label{fund}
\end{equation}
where $\tau=t_2-t_1$. The parameter $\varepsilon$ is determined by
demanding that $H_{int}(t_2,t_1)$  called the generalized
interaction operator must be so close to the solution of Eq.
(\ref{main}) in the limit $t_2\tend t_1$ that this equation has a
unique solution having the behavior (\ref{fund}) near the point
$t_2=t_1$. If $H_{int}(t_2,t_1)$ is specified, Eq. (\ref{main})
allows one to find the operator $\tilde S(t_2,t_1)$, and hence the
evolution operator. Thus Eq. (\ref{main})  can be regarded as an
equation of motion for states of a quantum system.  This
generalized dynamical equation  allows one to construct the
evolution operator by using the contributions from fundamental
processes as building blocks. In the case of Hamiltonian dynamics
the fundamental interaction is instantaneous. The generalized
interaction operator describing such an interaction is of the form
\begin{equation}
 H_{int}(t_2,t_1) = - 2i \delta(t_2-t_1)
 H_{I}(t_1) \label{loc}
\end{equation}
(the delta function $\delta(t_2-t_1)$ emphasizes that the
interaction is instantaneous). In this case Eq. (\ref{main}) is
equivalent to the Schr{\"o}dinger equation \cite{R.Kh.:1999} with
the operator $H_I(t)$ being an interaction Hamiltonian (in the
interaction picture). At the same time, Eq. (\ref{main}) permits
the generalization to the case where the fundamental interaction
in
 a quantum system is nonlocal in time, and hence
the dynamics is non-Hamiltonian \cite{R.Kh.:1999}.

Let us now briefly consider (for details see Ref.
\cite{R.Kh.:1999}) the mathematical assumptions on which the
formalism of the GQD is founded. As we already noted, in the
canonical formalism the assumption that the evolution of a quantum
system is governed by the Schr{\"o}dinger equation is used as a
dynamical postulate. On the other hand, since the operator
$V(t)\equiv U(t,0)$ is postulated to be unitary with the group
properties
$$V(t_2,t_1)=V(t_2)V(t_1),\quad V(0)=1,$$
from Stone's theorem it follows that, if this one-parameter group
is weekly continuous, i.e.,
\begin{eqnarray}\label{Stone}
\langle \psi_2|V(t_2)|\psi_1\rangle\tend\limits_{t_2\to t_1}
\langle \psi_2|V(t_1)|\psi_1\rangle
\end{eqnarray}
for any $|\psi_1\rangle$ and $|\psi_2\rangle$ belonging to the
Hilbert space, than it has a self-adjoint infinitesimal generator
$H$:
$$V(t)=\exp (-iHt),\quad idV(t)/dt=HV(t).$$
Identifying $H$ with the total Hamiltonian as usual, we get the
Schr{\"o}dinger equation. Thus the assumption that this equation
governs the dynamics of a system is equivalent to the assumption
that the evolution operator satisfies the condition (\ref{Stone}).
However, from the physical point of view, Eq. (\ref{Stone}) must
not be satisfied for any states belonging to the Hilbert space. In
fact, there are normalized vectors in the Hilbert space that
represent the states for which energy of a system is infinite.
From the point of view of the states with infinite energy
$|\psi_{in}\rangle$ any time interval $\delta t$ is infinite, and
hence the evolution operator $U(\delta t,0)\equiv 1+R(\delta t,
0)$ must be independent of $\delta t$, i.e., must be constant
\begin{eqnarray}
\langle \psi'_{in}|U(\delta t,0)|\psi_{in}\rangle=\langle
\psi'_{in}|\psi_{in}\rangle+\langle
\psi'_{in}|R(\infty,0)|\psi_{in}\rangle.\nonumber
\end{eqnarray}
This means that for the evolution operator to be weekly
continuous, $\langle\psi'_{in}|R(\infty,0)|\psi_{in}\rangle$ must
be zero. However, in quantum field theory, for example, this is
not the case because in this theory renormalization is necessary
to effectively cut off intermediate states with infinite energy.
On the other hand, the states $|\psi_{in}\rangle$ are not
physically realizable, and hence from the physical point of view
it is enough to require that the condition (\ref{Stone}) must be
satisfied only for the physically realizable states. The main idea
of the formalism of the GQD is that in describing the dynamics of
a system one need not to restrict oneself to the assumption that
the evolution operator is necessarily weekly continuous, and
therefore dynamics is necessarily governed by the Schr{\"o}dinger
equation. In Ref. \cite{R.Kh.:1999} it has been shown that the
current concepts of quantum physics generate the dynamical
principle in which the above assumption does not play any role. In
this way the GDE has been derived as the most general dynamical
equation. This equation allows one to construct the evolution
operator starting with the contributions from the processes with
infinitesimal duration times of interaction that are described by
the interaction operator $H_{int}(t_2,t_1)$, while the
Schr{\"o}dinger equation constructs the evolution operator
starting with the contributions from the processes with
instantaneous interaction that are described by the interaction
Hamiltonian. The GDE allows one to describe, in a consistent way,
the dynamics of a quantum system with a nonlocal in time
interaction. This equation is equivalent to the Schr{\"o}dinger
equation in the case when the interaction in a system is
instantaneous. In the case of nonlocal-in-time interactions the
dynamics is determined by the behavior of the interaction operator
$H_{int}(t_2,t_1)$ in the infinitesimal neighborhood of the point
$t_2=t_1$, i.e., such a nonlocality has no a definite scale. This
scale should depend on the problem we consider: Some duration time
of interaction may seem as infinitesimal from the point of view of
low energy physics, but at the same time it may seem as finite
from viewpoint of high energy physics. Thus the GDE allows one to
take into account, in a consistent way, that every theory with
which we deal is a low energy approximation to a more fundamental
one and provides a bridge between them. This is may be important
for describing the dynamics in the effective theory of nuclear
forces. In this case the duration times of interaction much above
the time scale of low energy nuclear physics but much below the
scale at which high energy degrees of freedom come into play
should be considered as "infinitesimal", and the operator
describing the processes with such duration times of interaction
can be used as an effective interaction operator.

By using Eq. (\ref{ev}), for $U(t,t_0)$, we can write
\begin{eqnarray}
U(t,t_0)&=&  {\bf 1} + \frac{i}{2\pi}
\int^\infty_{-\infty} dx\exp[-i(z-H_0)t]\label{evo}\\
&\times &(z-H_0)^{-1}T(z)(z-H_0)^{-1} \exp[i(z-H_0)t_0],\nonumber
\end{eqnarray}
 where $z=x+iy$, $y>0$, and $H_0$ is the free Hamiltonian.
 The
 operator $T(z)$ is defined by
\begin{eqnarray}
T(z) = i \int_{0}^{\infty} d\tau \exp(iz\tau)\tilde
T(\tau),\label{tz}
\end{eqnarray}
where $\tilde T(\tau)=\exp(-iH_0t_2)\tilde
S(t_2,t_1)\exp(iH_0t_1)$, and $\tau=t_2-t_1$.
 In terms of the
$T$-matrix defined by Eq. (\ref{tz}) the equation of motion
(\ref{main}) can be rewritten in the form \cite{R.Kh.:1999}
\begin{equation}
\frac{d \langle \psi_2|T(z)|\psi_1\rangle}{dz} = -  \sum
\limits_{n} \frac{\langle \psi_2|T(z)|n\rangle\langle
n|T(z)|\psi_1\rangle}{(z-E_n)^2},\label{difer}
\end{equation}
where $n$ stands for the entire set of discrete and continuous
variables that characterize the system in full, and $|n\rangle$
are the eigenvectors of $H_0$. As it follows from Eqs.
(\ref{fund}) and (\ref{tz}), the boundary condition on this
equation is of the form
\begin{eqnarray}
\langle\psi_2|T(z)|\psi_1\rangle \tend \limits_{|z|\tend\infty} \langle\psi_2|B(z)|\psi_1\rangle
+o\left(|z|^{-\beta}\right)\nonumber\\
=\langle\psi_2|B(z)|\psi_1\rangle
+O\left\{h(z)\right\},\label{T(z)to}\\
B(z) = i \int_{0}^{\infty} d\tau \exp(iz\tau) H_{int}^{(s)}(\tau),
\label{B(z)-Hint}
\end{eqnarray}
 where
$$H^{(s)}_{int}(t_2-t_1) = \exp(-iH_0t_2) H_{int}(t_2,t_1)
\exp(iH_0t_1)$$ is the interaction operator in the Schr{\"o}dinger
picture, and $h(z)$ is an arbitrary function satisfying the
condition $h(z)=o\left(|z|^{-\beta}\right)$, $|z|\to\infty$, with
$\beta=\varepsilon+1$. In the case of the Hamiltonian dynamics
when $H_{int}^{(s)}(\tau)=-2i\delta(\tau)H_I$, with $H_I$ being
the interaction Hamiltonian in the Schr{\"o}dinger picture,
 $B(z)=H_I$, and
the boundary condition (\ref{T(z)to}) takes the form
\begin{eqnarray}
T(z) \tend \limits_{|z|\tend\infty} H_I. \label{B(z)-H_I}
\end{eqnarray}
Equation (\ref{difer}) with this boundary condition is equivalent
to the Lippmann-Schwinger (LS) equation with the interaction
Hamiltonian $H_I$.
 By definition, the operator $B(z)$
 is the interaction operator in the energy
representation. This operator must be so close to the relevant
solution of Eq. (\ref{difer}) in the limit $|z|\to\infty$ that
this differential equation has a unique solution having the
asymptotic behavior (\ref{T(z)to}). For this the operator $B(z)$
must satisfy the condition
\begin{eqnarray}
\frac{d \langle \psi_2|B(z)|\psi_1\rangle}{dz} &=& - \sum
\limits_{n} \frac{\langle \psi_2|B(z)|n\rangle\langle n |B(z)|\psi_1\rangle}{(z-E_n)^2}\nonumber\\
&+&o(|z|^{-\beta-1}), \qquad |z|\to\infty. \label{diferB}
\end{eqnarray}

From Eq. (\ref{evo}) it follows that the evolution operator in the
Schr{\"o}dinger picture can be represented in the form
\begin{eqnarray}
U_s(t,0)=\frac{i}{2\pi}\int\limits_{-\infty}^{\infty}dx\exp(-izt)
G(z),\label{U-G}
\end{eqnarray}
where $z=x+iy$, $y>0$, and
\begin{eqnarray}
 G(z)=G_0(z)+G_0(z)T(z)G_0(z)\label{G-T}
\end{eqnarray}
with $G_0(z)=\frac{1}{z-H_0}.$ Being equivalent to representation
(\ref{ev}), Eqs. (\ref{U-G}) and (\ref{G-T}) express the principle
of the superposition of probability amplitudes. In the ordinary
quantum mechanics the similar equation plays an important role and
establishes the connection between the evolution operator and the
Green operator $G(z)$ which is defined by
\begin{eqnarray}
 G(z)=\frac{1}{z-H},\label{G}
\end{eqnarray}
with $H$ being the total Hamiltonian. Such a form of the Green
operator follows from the fact that in the Hamiltonian formalism
the evolution operator satisfies the Schr{\"o}dinger equation. In
the canonical formalism the $T$-matrix is defined by Eq.
(\ref{G-T}) starting with the Green operator of the form
(\ref{G}). In the formalism of the GQD the $T$-matrix plays a more
fundamental role. It is defined by Eq. (\ref{tz}), and in this
case the starting point is representation (\ref{ev}) with $\tilde
S(t_2,t_1)$ being the contribution to the evolution operator from
the processes in which the interaction begins at time $t_1$ and
ends at time $t_2$. In this case the operator $G(z)$ itself is
defined by Eq. (\ref{G-T}) via the $T$-matrix. This is a more
general definition of the Green operator, since representation
(\ref{ev}) is a consequence of the principle of the superposition
of probability amplitudes and must be valid in any case while the
evolution operator can be represented in the form (\ref{U-G}) with
the operator $G(z)$ given by Eq. (\ref{G}) only in the case where
the interaction in a system is instantaneous.

As has been shown in Ref. \cite{PRC:2002}, there is a one-to-one
correspondence between the character of the dynamics and the
large-momentum behavior of the $T$-matrix. If this behavior
satisfies the requirements of ordinary quantum mechanics, then the
interaction in a system is instantaneous and the dynamics is
Hamiltonian. In the case where this behavior is "bad", i.e., does
not meet the requirements of the ordinary quantum mechanics, the
interaction generating the dynamics of the system must necessarily
be nonlocal-in-time. Let us now illustrate this point by using the
model developed in Refs. \cite{R.Kh.:1999,PRC:2002} as a test
model demonstrating the possibility of the extension of quantum
dynamics provided by the GQD. This model describes the evolution
of the system of two identical nonrelativistic particles with the
mass $m$ whose interaction is separable, and hence the interaction
operator has the form
\begin{eqnarray}
 <{\bf p}_2| H_{int}^{(s)}(\tau)|{\bf p}_1>
 = \varphi^*({\bf p}_2)\varphi({\bf p}_1)f(\tau),\nonumber
\end{eqnarray}
where ${\bf p}$ is the relative momentum of the particles, and
$f(\tau)$ is some function of the duration time $\tau$ of the
interaction in the system.
 It is assumed that in the limit $|{\bf p}|\to\infty$
the form factor $\varphi({\bf p})$ behave as
\begin{equation}
\varphi({\bf p}) \sim |{\bf p}|^{-\alpha}, \quad {(|{\bf p}| \tend
\infty).}\nonumber
\end{equation}
In this case the general solution of Eq. (\ref{difer}) is
\cite{PRC:2002}
\begin{eqnarray}
\langle{\bf p}_2|T(z)|{\bf p}_1\rangle =\frac{ \varphi^* ({\bf
p}_2)\varphi({\bf p}_1)}{g_a^{-1} +(z-a) \int
\frac{d^3k}{(2\pi)^3} \frac {|\varphi({\bf
k})|^2}{(z-E_k)(a-E_k)}},\label{sol}
\end{eqnarray}
where $g_a=t(a)$, and $a\in(-\infty,0]$. In the case
$\alpha>\frac{1}{2}$, the amplitude $\langle{\bf p_2}|T(z)|{\bf
p_1}\rangle$ given by Eq. (\ref{sol}), tends to $\lambda
 \varphi^*({\bf p}_2)\varphi({\bf p}_1)$, where
$$\lambda=\left(g_a^{-1}+\int\frac{d^3k}{(2\pi)^3}
\frac{|\varphi({\bf k})|^2}{a-E_k}\right)^{-1}.$$ From Eqs.
(\ref{T(z)to}) and (\ref{B(z)-Hint}) it follows that in this case
the interaction operator $H_{int}^{(s)}(\tau)$ should be of the
form
$$<{\bf p}_2| H_{int}^{(s)}(\tau)|{\bf p}_1>
 = -2i\lambda\delta(\tau)\varphi^*({\bf p}_2)\varphi({\bf p}_1),
$$
and hence the dynamics of the system is Hamiltonian and is
governed by the Schr{\"o}dinger equation with the potential
$\langle{\bf p_2}|V|{\bf p_1}\rangle=\lambda
 \varphi^*({\bf p}_2)\varphi({\bf p}_1)$.
 In the case $\alpha<\frac{1}{2}$, the $T$-matrix (\ref{sol})
 tends to zero as $|z|\tend\infty$
 \begin{eqnarray}
\langle{\bf p_2}|T(z)|{\bf p_1}\rangle \tend \limits_{|z| \tend
\infty} \varphi^* ({\bf p}_2)\varphi({\bf p}_1)\nonumber\\
\times\left( b_1 (-z)^{\alpha-\frac{1}{2}}+ b_2 (-z)^{2
\alpha-1}\right) + o(|z|^{2 \alpha-1}),\nonumber
\end{eqnarray}
where $b_1 =- 4\pi \cos(\alpha \pi) m^{\alpha-\frac{3}{2}}$ and
$b_2= b_1 |a|^{\frac{1}{2}- \alpha} -b_1^2(\tilde M(a)+g_a^{-1})$
with
$$
\tilde M(a) = \int\frac{d^3k}{(2\pi)^3} \frac {|\varphi({\bf
k})|^2-
 |{\bf {k}}|^{-2\alpha}}
{a-E_k} .
$$
In this case the dynamics is non-Hamiltonian and, as it follows
from Eqs. (\ref{T(z)to}) and (\ref{B(z)-Hint}), is generated by
the nonlocal-in-time interaction operator
\begin{eqnarray} <{\bf p}_2| {H}^{(s)}_{int}(\tau)|{\bf p}_1>
& = &\varphi^* ({\bf p}_2)\varphi ({\bf p}_1)\nonumber\\
&\times&\left( a_1\tau^{-\alpha-\frac{1}{2}} + a_2 \tau^{-2
\alpha}\right), \label{less}
\end{eqnarray}
where $a_1= 4\pi i\cos(\alpha \pi) m^{\alpha-\frac{3}{2}} \Gamma
^{-1}(\frac{1}{2}-\alpha) \exp[i(-\frac{\alpha}{2}+ \frac{1}{4})
\pi]$, and $a_2$ is a free parameter of the theory. The solution
of Eq. (\ref{difer}) with the interaction operator (\ref{less}) is
of the form
\begin{eqnarray}
\langle{\bf p}_2| T(z)|{\bf p}_1\rangle =\frac{b_1^2\varphi^*
({\bf p}_2)\varphi ({\bf
p}_1)}{-b_2+b_1(-z)^{\frac{1}{2}-\alpha}-\tilde
M(z)b_1^2}.\nonumber
\end{eqnarray}

\section{The pionless theory of nuclear forces}

The Weinberg program for low energy nucleon physics employs the
analysis of time-ordered diagrams for the $2N$ $T$-matrix in ChPT
to derive a $NN$ potential and then to use it in the LS equation
for constructing the full $NN$ $T$-matrix. Obviously the starting
point for this program is the assumption that in the
nonrelativistic limit ChPT leads to low energy nucleon dynamics
which is Hamiltonian and is governed by the Schr{\"o}dinger
equation. However, the fact that the chiral potentials constructed
in this way are singular and lead to UV divergences means that
this is not the case. At the same time, the GQD allows one to
analyze the  predictions of ChPT without making $a$ $priori$
assumptions about the character of low energy nucleon dynamics:
This character should results from the analysis. Let us consider,
for  example, the low energy predictions of ChPT for the $2N$
system in the ${}^1S_0$ channel. At very low energy even the pion
field can be integrated out, and the diagrams of the ChPT take the
form of the diagrams being produced by the effective Lagrangian
containing only contact interactions among nucleons and
derivatives thereof \cite{Kaplan2}. From the analysis of
time-ordered diagrams of this theory it follows that the $2N$
$T$-matrix in the ${}^1S_0$ channel must be of the form
\begin{equation}
\langle {\bf p_2}|T(z)|{\bf p_1}\rangle = \sum
\limits_{n,m=0}^\infty p_2^{2n}p_1^{2m}t_{nm}(z),\\ \label{anW}
|{\bf p}_1|<\Lambda, \quad|{\bf p}_2|<\Lambda,\nonumber
\end{equation}
where ${\bf p}_i$ is relative momentum of nucleons, and the
functions $t_{nm}(z)$ are of order $O\{(Q/\Lambda)^{2n+2m}\}$. On
the other hand, the $2N$ $T$-matrix must satisfy the GDE of the
form (\ref{difer}). From Eq. (\ref{anW}) it follows that for
$T(z,{\bf p_2} ,{\bf p_1} )\equiv\langle {\bf p_2}|T(z)|{\bf
p_1}\rangle$, we can write
\begin{eqnarray}\label{asym_T}
T(z,{\bf p_2},{\bf p_1})=\sum\limits_{n=0}^\infty p_2^{2n}{\cal
T}_n(z,{\bf p}_1),\quad |{\bf p_2}|<\Lambda,\nonumber\\
T(z,{\bf p_2},{\bf p_1})=\sum\limits_{n=0}^\infty
p_1^{2n}\tilde{\cal T}_n(z,{\bf p}_2),\quad |{\bf
p_1}|<\Lambda,\nonumber\\
{\cal T}_n(z,{\bf p})=\sum\limits_{m=0}^\infty
p^{2m}t_{nm}(z), \quad |{\bf p}|<\Lambda,\nonumber\\
\tilde{\cal T}_n(z,{\bf p} )=\sum\limits_{m=0}^\infty p^{2m}
t_{mn}(z), \quad |{\bf p}|<\Lambda.
\end{eqnarray}
By using these expansions, Eq. (\ref{difer}) for the $2N$
$T$-matrix can be written in the form
\begin{eqnarray}\label{dif_0}
\frac{dT(z,{\bf p_2} ,{\bf p_1} )}{dz}&=&\tilde{\cal T}_0(z,{\bf
p_2} ){\cal T}_0(z,{\bf
p_1} )\frac{dM_0(z)}{dz}\nonumber\\
&+&{\cal F}(z,{\bf p_2},{\bf p_1}),
\end{eqnarray}
with
$$M_0(z)=zm\int\frac{d^3k}{(2\pi)^3}\frac{1}{(zm-k^2)E_k}=\frac{m}{4\pi}\sqrt{-zm},$$
\begin{equation}\label{F=}
{\cal F}(z,{\bf p_2},{\bf
p_1})=-\int\frac{d^3k}{(2\pi)^3}\frac{F_1(z,{\bf p_2},{\bf
p_1},{\bf k}/\Lambda)}{(z-E_k)^2},
\end{equation}
\begin{eqnarray}
{\cal F}_1(z,{\bf p_2},{\bf p_1},{\bf k}/\Lambda)&=&T(z,{\bf
p_2},{\bf k})T(z,{\bf k},{\bf p_1}) \nonumber\\&-&\tilde{\cal
T}_0(z,{\bf p}_2){\cal T}_0(z,{\bf p}_1).\nonumber
\end{eqnarray}
The function ${\cal F}(z,{\bf p_2},{\bf p_1})$ is of order
$O\{Q/\Lambda\}$. This is because, Eq. (\ref{F=}) can be rewritten
in the form
\begin{eqnarray}
{\cal F}(z,{\bf p_2},{\bf
p_1})=-\frac{m}{\Lambda}\int\frac{d^3\tilde
k}{(2\pi)^3}\frac{F_1(z,{\bf p_2},{\bf p_1},\tilde{\bf
k})}{(mz/\Lambda^2-\tilde k^2)^2},\nonumber
\end{eqnarray}
where $\tilde{k}=k/\Lambda$ and $F_1(z,{\bf p_2},{\bf
p_1},\tilde{\bf k})=o\{\tilde k^2\},$ $\tilde k\to 0$, due to the
above substraction. This allows one to construct the $2N$
$T$-matrix in the spirit of the effective theory by expanding it
in powers of $Q/\Lambda$
\begin{eqnarray}
T(z,{\bf p_2} ,{\bf p_1} )=T^{(2{\cal N})}(z,{\bf p_2} ,{\bf p_1}
)+O\left\{(Q/\Lambda)^{2{\cal N}+1}\right\},\nonumber
\end{eqnarray}
where $T^{({2\cal N})}(z,{\bf p_2},{\bf p_1})$ is the sum of the
first ${\cal N}$ terms in the expansion. The function $T^{(2{\cal
N})}(z,{\bf p_2},{\bf p_1})$ must satisfy Eq. (\ref{dif_0}) with
an accuracy of  order $O\left\{(Q/\Lambda)^{2{\cal N}+1}\right\}$
\begin{eqnarray}\label{dif_0up}
\frac{dT^{(2{\cal N})}(z,{\bf p_2} ,{\bf p_1}
)}{dz}=\Big[\tilde{\cal T}^{(2{\cal N})}_0(z,{\bf p_2} ){\cal
T}^{(2{\cal N})}_0(z,{\bf p_1}
)\nonumber\\\times\frac{dM_0(z)}{dz} +{\cal F}^{(2{\cal
N})}(z,{\bf p_2} ,{\bf p_1}
)\Big]
\left(1+O\left\{(Q/\Lambda)^{2{\cal
N}+1}\right\}\right),\nonumber
\end{eqnarray}
where
\begin{eqnarray}
{\cal F}^{(2{\cal N})}(z,{\bf p_2} ,{\bf p_1}
)=-\int\frac{d^3k}{(2\pi)^3}\frac{1}{(z-E_k)^{2}} \Big(T^{(2{\cal
N})}(z,{\bf
p_2} ,{\bf k} )\nonumber\\\times T^{(2{\cal N})}(z,{\bf k} ,{\bf p_1})
-\tilde{\cal T}^{(2{\cal N})}_0(z,{\bf p}_2){\cal T}^{(2{\cal
N})}_0(z,{\bf p}_1)\Big).\nonumber
\end{eqnarray}
Taking into account that $F(z,{\bf p_2},{\bf p_1})$ is of order
$O\{Q/\Lambda\}$, Eq. (\ref{dif_0up}) can be rewritten in the form
\begin{eqnarray}\label{dif_again}
\frac{dT^{(2{\cal N})}(z,{\bf p_2} ,{\bf p_1}
)}{dz}&=&\Big[\tilde{\cal T}^{(2{\cal N})}_0(z,{\bf p_2} ){\cal
T}^{(2{\cal N})}_0(z,{\bf p_1}
)\nonumber\\\times\frac{dM_0(z)}{dz} &+&{\cal F}^{(2({\cal
N}-1))}(z,{\bf p_2} ,{\bf p_1}
)\Big]\nonumber\\&\times&\left(1+O\left\{(Q/\Lambda)^{2{\cal
N}+1}\right\}\right).
\end{eqnarray}
This equation allows one to construct the $2N$ $T$-matrix order by
order.

At leading order Eq. (\ref{dif_again}) takes the form
\begin{eqnarray}\label{dif_at lead}
\frac{dT^{(0)}(z,{\bf p_2} ,{\bf p_1} )}{dz}&=&\tilde{\cal
T}^{(0)}_0(z,{\bf p_2} ){\cal T}^{(0)}_0(z,{\bf p_1}
)\nonumber\\&\times&\frac{dM_0(z)}{dz}\Big(1+O\{Q/\Lambda\}\Big).
\end{eqnarray}
By using this equation and the above expansion of $T(z,{\bf
p}_2,{\bf p}_1)$, for $t_{00}(z)$ at leading order (LO) we get the
equation
\begin{eqnarray}
\frac{dt^{(0)}_{00}(z)}{dz}=\left(t^{(0)}_{00}(z)\right)^2\frac{dM_0(z)}{dz}\Big(1+O\{Q/\Lambda\}\Big).\nonumber
\end{eqnarray}
The solution of this equation is
\begin{equation}
t^{(0)}_{00}(z)=\left(C^{-1}_0-M_0(z)\right)^{-1}\Big(1+O\{Q/\Lambda\}\Big),\nonumber
\end{equation}
where $C_0$ is some constant. Correspondingly, for ${\cal
T}^{(0)}_0(z,{\bf p} )$ we have the equation
\begin{eqnarray}
\frac{d{\cal T}^{(0)}_0(z,{\bf p} )}{dz}=t^{(0)}_{00}(z){\cal
T}^{(0)}_0(z,{\bf p}
)\frac{dM_0(z)}{dz}
\Big(1+O\{Q/\Lambda\}\Big).\nonumber
\end{eqnarray}
whose solution is
\begin{equation}\label{sol_cal T}
{\cal T}^{(0)}_0(z,{\bf p} )=t^{(0)}_{00}(z)\Psi_0({\bf
p})\Big(1+O\{Q/\Lambda\}\Big),
\end{equation}
where $\Psi_0({\bf p})$ is some function that for $|{\bf
p}|<\Lambda$ can be expanded as
\begin{equation}\label{Psi_0}
  \Psi_0({\bf p})=\sum\limits_{n=0}^\infty c_{2n}p^{2n},\quad |{\bf
  p}|<\Lambda,
\end{equation}
with $c_0=1$. In the same way, for $\tilde{\cal T}^{(0)}_0(z,{\bf
p} )$, we get
\begin{equation}\label{sol_tilde T}
\tilde{\cal T}^{(0)}_0(z,{\bf p}
)=t^{(0)}_{00}(z)\tilde\Psi_0({\bf
p})\left(1+O\{Q/\Lambda\}\right),
\end{equation}
where the function $\tilde\Psi_0(p )$ satisfies the condition
\begin{eqnarray}\label{tilde Psi_0}
 \tilde \Psi_0({\bf p})=\sum\limits_{n=0}^\infty c'_{2n}p^{2n},\quad |{\bf
  p}|<\Lambda.\nonumber
\end{eqnarray}
Now, substituting $\tilde{\cal T}^{(0)}_0(z,{\bf p} )$ and ${\cal
T}^{(0)}_0(z,{\bf p} )$  into Eq. (\ref{dif_at lead}), we can
obtain the amplitude $T(z,{\bf p_2},{\bf p_1})$ at LO
\begin{eqnarray}\label{sol_at lead}
T^{(0)}(z,{\bf p_2} ,{\bf p_1} )&=&\Big[\tilde\Psi_0({\bf
p_2} )\Psi_0({\bf p_1})\left(C^{-1}_0-M_0(z)\right)^{-1}\nonumber\\
&+&\Phi_0({\bf p_2},{\bf p_1})\Big] \left(1+O\{Q/\Lambda\}\right),
\end{eqnarray}
where $\Phi_0({\bf p_2} ,{\bf p_1} )$ is an arbitrary function
satisfying the following conditions
\begin{eqnarray}
\Phi_0({\bf p_1},{\bf p_2})=\sum \limits_{n=1}^\infty
p_1^{2n}\varphi_{2n}({\bf p_2}),\quad |{\bf{p_1}}|<\Lambda,\nonumber\\
\Phi_0({\bf p_2},{\bf p_1})=\sum \limits_{n=1}^\infty
p_1^{2n}\tilde\varphi_{2n}({\bf p_2}),\quad |{\bf{p_1}}|<\Lambda,\nonumber\\
\varphi_{2n}({\bf p})=\sum \limits_{m=1}^\infty {\cal C}_{2n,2m}
p^{2m}, \quad |{\bf p}|<\Lambda,\nonumber\\
\tilde\varphi_{2n}({\bf p})=\sum \limits_{m=1}^\infty {\cal
C}_{2m,2n} p^{2m}, \quad |{\bf p}|<\Lambda.\nonumber
\end{eqnarray}
These conditions result from the fact that, for the function
$\Phi_0({\bf p_2} ,{\bf p_1} )$  to be consistent with the
structure of the $2N$ $T$-matrix shown in Eq. (\ref{anW}), it
should be of the form
\begin{eqnarray}\label{Phi}
\Phi_0({\bf p_2},{\bf p_1})=\sum \limits_{n,m=0}^\infty {\cal
C}_{2n,2m} p_2^{2n}p_1^{2m},\nonumber\\ \quad |{\bf
p_1}|<\Lambda,|{\bf p_2}|<\Lambda.
\end{eqnarray}
On the other hand, ${\cal C}_{0,2m}$ and ${\cal C}_{2n,0}$ should
be zero because if this is not the case the terms ${\cal C}_{0,0}$
and ${\cal C}_{0,2n}p_1^{2n}$ have to be included into
$t_{00}^{(0)}(z)$ and $T_0^{(2)}(z,{\bf p}_1)$, respectively.
 As
has been shown in Ref. \cite{R.Kh.:1999}, for the solution of Eq.
(\ref{difer}) to be unitary, it must satisfy the condition
\begin{eqnarray}
  \langle {\bf p}_2|T^+(z)|{\bf p}_1\rangle=\langle {\bf p}_2|T(z)|{\bf
  p}_1\rangle,\quad z\in(-\infty,0).\nonumber
\end{eqnarray}
From this and Eq. (\ref{sol_at lead}) it follows that
$\tilde\Psi_0({\bf p})=\Psi_0^*({\bf p})$,
$\tilde\varphi_{2n}({\bf p})=\varphi^*_{2n}({\bf p})$, ${\cal
C}_{2n,2m}={\cal C}_{2m,2n}^*$, and $\Phi_0^*({\bf p_2},{\bf
p_1})=\Phi_0({\bf p_1},{\bf p_2})$.

The fact that Eq. (\ref{sol_at lead}) represents the $2N$
$T$-matrix at leading order means that at extreme low energies the
true $2N$ $T$-matrix in the ${}^1S_0$ channel is described by this
equation. In other words, at $z=0$ the true $2N$ $T$-matrix should
coincide with $T^{(0)}(0,{\bf p}_2,{\bf p}_1)$ given by Eq.
(\ref{sol_at lead}), and hence the higher order corrections must
not change $\langle {\bf p}_2|T(z)|{\bf p}_1\rangle$ at $z=0$.
This puts the boundary condition on $T^{(2{\cal N})}(z,{\bf
p}_2,{\bf p}_1)$
\begin{eqnarray}\label{boundcond}
T^{(2{\cal N})}(0,{\bf p}_2,{\bf p}_1)=\Psi_0^*({\bf
p}_2)\Psi_0({\bf p}_1)C_0+\Phi_0({\bf p_2},{\bf p_1}).
\end{eqnarray}
By using Eq. (\ref{dif_again}) with this boundary condition, one
can obtain the $2N$ $T$-matrix at any order. However, the
functions $\Psi_0({\bf p})$ and $\Phi_0({\bf p_2},{\bf p_1})$ that
occur the boundary condition (\ref{boundcond}) are not known
exactly because such a knowledge implies that we know the details
of the underlying dynamics (the relevant domain of the definition
of the functions $\Psi_0({\bf p})$ and $\Phi_0({\bf p_2},{\bf
p_1})$ spreads to high energies). As we show below, for obtaining
$T(z,{\bf p}_2,{\bf p}_1)$ with an accuracy of order
$O\{(Q/\Lambda)^{2{\cal N}+1}\}$ it is sufficient to know a few
constants $c_{2n}$ and ${\cal C}_{n,m}$ that appear in the
expansions (\ref{Psi_0}) and (\ref{Phi}), and a few parameters
characterizing integral properties of the functions $\Psi_0({\bf
p})$ and $\Phi_0({\bf p_2},{\bf p_1})$. This allows one to
describe low energy nucleon dynamics in terms of a few low energy
constants (LEC's) without any knowledge of the details of the
dynamics at high energies.

For describing low energy nucleon dynamics at leading order it is
sufficient to restrict oneself to the first term in the expansion
(\ref{Psi_0}) of the function $\Psi_0({\bf p})$, while the
function $\Phi_0({\bf p_2},{\bf p_1})$ can be omitted because
${\cal C}_{0,0}=0$. In this case the LO $2N$ $T$-matrix takes the
form
\begin{equation}
\langle{\bf p}_2| T^{(0)}(z)|{\bf p}_1\rangle
=\left(C_0^{-1}-\frac{m}{4\pi}\sqrt{-zm}\right)^{-1},\label{T^(0)}
\end{equation}
where the low energy constant (LEC) $C_0$ parametrizes the effects
of high energy physics on low energy nucleon dynamics. This
constant can be obtained by fitting to the $2N$ scattering data.
Since the scattering amplitude $A(p)=-\langle{\bf p}|
T^{(0)}(E_p+i0)|{\bf p}\rangle$ at $p=0$ should be described by
the LO $2N$ $T$-matrix, the constant $C_0$ is related to the
scattering length $a$ as $C_0=4\pi a/m$. Note that the standard
EFT approach yields the same expression for the LO $T$-matrix
\cite{{EFT3},{Kaplan},{vKolck}}.  It is remarkable that the
$T$-matrix (\ref{T^(0)}), which in the EFT approach is obtained by
performing regularization and renormalization of the solution of
the Schr{\"o}dinger (LS) equation, is not a solution of this
equation. At the same time, it is a solution of Eq. (\ref{difer})
with the interaction operator  \cite{PRC:2002}
\begin{equation}
\langle{\bf p}_2| B^{(0)}(z)|{\bf p}_1\rangle
=-\frac{4\pi}{m\sqrt{-zm}}+\frac{16\pi^2}{C_0m^2zm}. \label{B^0}
\end{equation}
The corresponding generalized interaction operator
$H_{int}^{(s)}(\tau)$ is of the form
\begin{equation}
\langle{\bf p}_2| H_{int}^{(s)}(\tau)|{\bf p}_1\rangle
=\Big(\frac{2\pi}{m}\Big)^{3/2}\frac{i-1}{\pi}\left(\tau^{-1/2}+\theta\right),\nonumber
\end{equation}
where $\theta=(\frac{2\pi}{m})^{3/2}C_0^{-1}(1+i)$. This operator
is nonlocal in time and this is the only reason why in this case
the GDE cannot be reduced to the LS equation, and hence the
$T$-matrix is not its solution.

At the same time, for using $T^{(0)}(z,{\bf p}_2,{\bf p}_1)$ in
Eq. (\ref{dif_again}) (more precisely the first term ${\cal
T}_0^{(0)}(z,{\bf p})$ in its momentum expansion (\ref{asym_T})),
one cannot restrict oneself to the $T$-matrix shown in Eq.
(\ref{T^(0)}). This is because in this case the LO $T$-matrix is
used for obtaining higher order corrections to $T(z,{\bf p}_2,{\bf
p}_1)$, and hence must contain a more detailed information than
the $T$-matrix responsible for describing nucleon dynamics at
leading order. Thus, while by definition $T(z,{\bf p}_2,{\bf
p}_1)$ is equivalent  to $\langle{\bf p}_2| T(z)|{\bf
p}_1\rangle$, the amplitudes $T^{(2{\cal N})}(z,{\bf p}_2,{\bf
p}_1)$ that are used in Eq. (\ref{dif_again}) and $\langle{\bf
p}_2| T^{(2{\cal N})}(z)|{\bf p}_1\rangle$ that control the
nucleon dynamics with an accuracy of order $O\{(Q/\Lambda)^{2{\cal
N}+1}\}$ represent different approaches to the $2N$ $T$-matrix: In
contrast with $\langle{\bf p}_2| T^{(2{\cal N})}(z)|{\bf
p}_1\rangle$ the amplitude $T^{(2{\cal N})}(z,{\bf p}_2,{\bf
p}_1)$ must contain the information that allows one not only to
describe low energy nucleon dynamics at leading order but also to
obtain the higher order corrections. At LO all the  needed
information is contained in the functions $\psi_0({\bf p})$ and
$\Phi_0({\bf p_2},{\bf p_1})$ that appear in Eq. ({\ref{sol_at
lead}). In principle they could be derived from QCD. However, for
describing low energy nucleon dynamics at LO one need not to know
these functions exactly. It is sufficient to know the first terms
in their low energy expansions. This manifests itself in the fact
that the $T$-matrix describing low energy nucleon dynamics at LO
is momentum independent. At the same time, as we show below, for
obtaining higher order corrections a better knowledge of the
properties of the functions $\psi_0({\bf p})$ and $\Phi_0({\bf
p_2},{\bf p_1})$ is needed.

At next to leading order $(NLO)$ Eq. (\ref{dif_again}) reads
\begin{eqnarray}\label{difNLO}
\frac{dT^{(2)}(z,{\bf p_2} ,{\bf p_1} )}{dz}=\Big[\tilde{\cal
T}^{(2)}_0(z,{\bf p_2} ){\cal T}^{(2)}_0(z,{\bf p_1}
)\nonumber\\\times\frac{dM_0(z)}{dz}+{\cal F}^{(0)}(z,{\bf p_2}
,{\bf p_1})\Big]\left(1+O\{Q^3/\Lambda^3\}\right),
\end{eqnarray}
and hence $t_{00}^{(2)}(z)$ satisfies the equation
\begin{widetext}
\begin{eqnarray}
\frac{dt_{00}^{(2)}(z)}{dz} =
\left[\left(t_{00}^{(2)}(z)\right)^2\frac{dM_0(z)}{dz}+\left(t_{00}^{(0)}(z)\right)^2\frac{dM(z)}{dz}\right]
\left(1+O\left\{Q^3/\Lambda^3\right\}\right),\label{d1N}
\end{eqnarray}
where $M(z) =zm\int \frac{d^3 k}{(2\pi)^3} \frac{F_1(k/\Lambda)}
{(zm-k^2)E_k}$ with $F_1(k/\Lambda)=|\Psi_0({\bf k})|^2-1$. The
function $M(z)$ can be expanded as
\begin{eqnarray}
M(z) =
zm\int\frac{d^3k}{(2\pi)^3}\frac{F_1(k/\Lambda)}{(zm-k^2)E_k}
=(zm)^2\int\frac{d^3k}{(2\pi)^3}\frac{F_1(k/\Lambda)
}{(zm-k^2)E_kk^2}-zm\cdot
m\int\frac{d^3k}{(2\pi)^3}\frac{F_1(k/\Lambda)}{k^4}\nonumber\\
=\ldots= \sum
\limits_{n=1}^{\infty}\left(\left(zm\right)^nM_0(z)\sum
\limits_{i=0}^{n}c^*_{2i}c_{2(n-i)}-J_{2n}(zm)^{n}\right) =\sum
\limits_{n=1}^{\infty}M_{2n}(z),\label{Mnz}
\end{eqnarray}
\end{widetext}
with
\begin{equation} M_{2n}(z)=\left(zm\right)^n\Big[M_0(z)\sum
\limits_{i=0}^{n}c^*_{2i}c_{2(n-i)}-J_{2n}\Big],
\end{equation}
where
\begin{eqnarray}\label{J_2j}
J_{2n} = m\int\frac{
d^3k}{(2\pi)^3}\frac{F_n(k/\Lambda)}{k^{2(n+1)}}\nonumber\\
=m\Lambda^{-2n+1}\int\frac{d^3\tilde{k}}{(2\pi)^3}
\frac{F_n(\tilde{k})}{\tilde{k}^{2(n+1)}}.
\end{eqnarray}
\begin{eqnarray}
F_n(k/\Lambda)=|\Psi_0({\bf k})|^2-\sum \limits_{i=0}^{n-1}\sum
\limits_{j=0}^{i}c^*_{2j}c_{2(i-j)}k^{2i},\nonumber
\end{eqnarray}
and $\tilde k=k/\Lambda$. From this it follows that the function
$M_{2n}(z)$ is of order $O\{Q^{2n}/\Lambda^{2n}\}$, and hence in
Eq. (\ref{d1N}) it is sufficient to keep only the first term in
the expansion (\ref{Mnz}). Thus Eq. (\ref{d1N}) can be rewritten
in the form
\begin{eqnarray}
\frac{dt_{00}^{(2)}(z)}{dz} &=&
\left[\left(t_{00}^{(2)}(z)\right)^2\frac{dM_0(z)}{dz}+\left(t_{00}^{(0)}(z)\right)^2\frac{dM_2(z)}{dz}\right]\nonumber\\
&\times&\left(1+O\left\{Q^3/\Lambda^3\right\}\right).\label{t_00^2}
\end{eqnarray}
The boundary condition for this equation should be as follows
\begin{equation}\label{C-0}
  t_{00}^{(2N)}(0)=C_0.
\end{equation}
This is because at $z=0$ the exact $2N$ $T$-matrix  should
coincide with the LO $T$-matrix and this puts the boundary
condition on $T^{(2{\cal N})}(z,{\bf p}_2,{\bf p}_1)$. The
solution of Eq. (\ref{t_00^2}) with this boundary condition is
\begin{eqnarray}
 t_{00}^{(2)}(z)=\left[C_0^{-1}-M_0(z)-M_2(z)\right]^{-1}.\nonumber
\end{eqnarray}
Now we can obtain the amplitude ${\cal T}_0(z,{\bf p})$ at NLO
that, as it follows from Eqs. (\ref{sol_cal T}), (\ref{sol_at
lead}), and (\ref{difNLO}), should satisfy the equation
\begin{eqnarray}
\frac{d{\cal T}_0^{(2)}(z,{\bf p})}{dz} =
\Big[t_{00}^{(2)}(z){\cal T}_0^{(2)}(z,{\bf p})\frac{dM_0(z)}{dz}\nonumber\\
+\left(t_{00}^{(0)}(z)\right)^2\frac{dM_2(z)}{dz}\Psi_0({\bf
p})+t_{00}^{(0)}(z)\frac{d{\cal M}(z,{\bf
p})}{dz}\Big]\nonumber\\
\times\left(1+O\left\{Q^3/\Lambda^3\right\}\right),\label{T_0^2}
\end{eqnarray}
where
\begin{eqnarray}
{\cal M}(z,{\bf p})=zm\int\frac{d^3k}{(2\pi)^3}\frac{\Psi_0({\bf
k})\Phi_0({\bf k},{\bf p})}{(zm-k^2)E_k}.\nonumber
\end{eqnarray}
 The function ${\cal M}(z,{\bf p})$ can be expanded as
\begin{eqnarray}
{\cal M}(z,{\bf p})={\cal M}_2(z,{\bf
p})\left(1+O\left\{Q^3/\Lambda^3\right\}\right)\nonumber\\
=\sum\limits_{n=1}^{\cal N}{\cal M}_{2n}(z,{\bf
p})\left(1+O\left\{(Q/\Lambda)^{2{\cal
N}+1}\right\}\right)\nonumber
\end{eqnarray}
with
\begin{eqnarray}
{\cal M}_{2n}(z,{\bf p})&=&(zm)^n
M_0(z)\sum\limits_{i=0}^{n-1}c_{2i}\varphi_{2(n-i)}({\bf
p})\nonumber\\&-&(zm)^n\alpha_{2n}({\bf p})\nonumber
\end{eqnarray}
where
\begin{eqnarray}
\alpha_{2n}({\bf
p})=m\int\frac{d^3k}{(2\pi)^3}k^{-2(n+1)}\Big[\Psi_0({\bf
k})\Phi_0({\bf k},{\bf p})
\nonumber\\
-\sum\limits_{j=1}^{n-1}k^{2j}\sum\limits_{i=0}^{j-1}c_{2i}\varphi_{2(j-i)}({\bf
p})\Big]. \nonumber
\end{eqnarray}
 This allows one to rewrite Eq. (\ref{T_0^2}) in the form
\begin{eqnarray}
\frac{d{\cal T}_0^{(2)}(z,{\bf p})}{dz} =
\Big[t_{00}^{(2)}(z){\cal T}_0^{(2)}(z,{\bf p})\frac{dM_0(z)}{dz}\nonumber\\
+\left(t_{00}^{(0)}(z)\right)^2\Psi_0({\bf
p})\frac{dM_2(z)}{dz}+t_{00}^{(0)}(z)\frac{d{\cal M}_2(z,{\bf p})}{dz}\Big]\nonumber\\
\times\left(1+O\left\{Q^3/\Lambda^3\right\}\right),\label{rT_0^2}
\end{eqnarray}
From Eq. (\ref{boundcond}) it follows that at $z=0$
\begin{eqnarray}\label{T(0,p)}
{\cal T}_0^{(2N)}(0,{\bf
p})=t_{00}^{(0)}(0)\Psi_0({\bf p})=C_0\Psi_0({\bf
p}).
\end{eqnarray}
 Solving Eq. (\ref{rT_0^2}) with this boundary condition
yields
\begin{eqnarray}\label{calt2}
{\cal T}_0^{(2)}(z,{\bf p})=\frac{\Psi_0({\bf p})+{\cal
M}_2(z,{\bf
p})}{C_0^{-1}-M_0(z)-M_2(z)}\nonumber\\
\times\left(1+O\left\{Q^3/\Lambda^3\right\}\right).
\end{eqnarray}
In the same way, for $\tilde{\cal T}_0^{(2)}(z,{\bf p})$, we get
\begin{eqnarray}\label{tilT2}
\tilde{\cal T}_0^{(2)}(z,{\bf p})=\frac{\Psi_0^*({\bf p})+\tilde
{\cal M}_2(z,{\bf p})}{C_0^{-1}-M_0(z)-M_2(z)}\nonumber\\
\times\left(1+O\left\{Q^3/\Lambda^3\right\}\right),
\end{eqnarray}
where $\tilde{\cal M}_2(z,{\bf p})={\cal M}_2^*(z^*,{\bf p})$.
Substituting these expressions into Eq. (\ref{difNLO}), we get the
equation
\begin{widetext}
\begin{eqnarray}\label{dT2}
\frac{d T^{(2)}(z,{\bf p}_2,{\bf p}_1)}{dz} = \Big[\Psi_0^*({\bf
p_2})\Psi_0({\bf
p_1})\left(t_{00}^{(0)}(z)\right)^2\left(\frac{dM_0(z)}{dz}+\frac{dM_2(z)}{dz}\right)\nonumber\\
+\left(\Psi_0^*({\bf p_2}){\cal M}_2(z,{\bf p_1})+\tilde {\cal
M}_2(z,{\bf
p}_2)\Psi_0({\bf p_1})\right)\left(t_{00}^{(2)}(z)\right)^2\frac{dM_0(z)}{dz}\nonumber\\
+t_{00}^{(0)}(z)\left(\Psi_0^*({\bf p_2})\frac{d{\cal M}_2(z,{\bf
p_1})}{dz}+\Psi_0({\bf p_1})\frac{d\tilde {\cal M}_2(z,{\bf
p}_2)}{dz}\right)\Big] \Big(1+O\left\{Q^3/\Lambda^3\right\}\Big).
\end{eqnarray}
Solving this equation with the boundary condition
(\ref{boundcond}) yields
\begin{eqnarray}\label{T^2(z,2,1)}
T^{(2)}(z,{\bf p}_2,{\bf p}_1)=\Big[\frac{\Psi_0^*({\bf
p_2})\Psi_0({\bf p_1})+\Psi_0^*({\bf p_2}){\cal M}_2(z,{\bf
p}_1)+\tilde {\cal M}_2(z,{\bf p}_2)\Psi_0({\bf
p_1})}{C_0^{-1}-M_0(z)-M_2(z)}+\Phi_0({\bf p}_2,{\bf
p}_1)\Big]\Big(1+O\left\{Q^3/\Lambda^3\right\}\Big).
\end{eqnarray}
 \end{widetext}
In describing low energy nucleon dynamics at NLO the terms
containing the functions $\Phi_0({\bf p}_2,{\bf p}_1)$,
$\varphi_2({\bf p})$ and $\alpha_2({\bf p})$ can be omitted, and
hence, for $\langle {\bf p}_2|T^{(2)}(z)|{\bf p}_1\rangle$, we get
\begin{eqnarray}\label{T2}
\langle {\bf p}_2|T^{(2)}(z)|{\bf p}_1\rangle =
\frac{\Psi_0^*({\bf p_2})\Psi_0({\bf
p_1})}{C_0^{-1}-M_0(z)-M_2(z)}\nonumber\\
\times\left(1+O\left\{Q^3/\Lambda^3\right\}\right).
\end{eqnarray}
For momenta below $\Lambda$ it is sufficient to keep the first two
terms in the expansion (\ref{Psi_0}), and the NLO $T$-matrix can
be represented in the form
\begin{eqnarray}\label{T^2}
& &\langle {\bf p}_2|T^{(2)}(z)|{\bf p}_1\rangle \nonumber\\&=&
\frac{(1+c_2^*p_2^2)(1+c_2p_1^2)}{C_0^{-1}-M_0(z)\Big(1+2\textsf{Re}c_2zm\Big)+{\cal J}_2zm}\nonumber\\
&\times&\left(1+O\left\{Q^3/\Lambda^3\right\}\right), \quad |{\bf
p}_1|<\Lambda,\quad |{\bf p}_2|<\Lambda.
\end{eqnarray}
 Thus, at NLO the effects of the
high energy physics that in the full theory are described by the
function $\Psi_0({\bf p})$ are parametrized by the two LEC's $c_2$
and ${\cal J}_2$. The constant $c_2$ determines the second term in
the expansion (\ref{Psi_0}) of the function $\Psi_0({\bf p})$,
while ${\cal J}_2$ characterizes its integral properties shown in
Eq. (\ref{J_2j}). Really for given $c_2$ and ${\cal J}_2$ Eq.
(\ref{T2}) represents the set (we will denote this set $\Omega_2$)
of the solutions of the GDE that coincides with the true $2N$
$T$-matrix with an accuracy of order $O\{Q^3/\Lambda^3\}$ provided
that $c_2$ and ${\cal J}_2$ are the LEC's chosen by nature.
Equation (\ref{sol_at lead}) also represents a solution set of the
GDE. The LO $T$-matrix given by Eq. (\ref{T^(0)}) belongs to this
set, and hence can be used for describing nucleon dynamics at
leading order. In contrast, Eq. (\ref{T^2}) does not represent the
$T$-matrix belonging to the set $\Omega_2$ because it does not
represent the solution of the GDE at all, and can describe the LO
$2N$ $T$-matrix only for momenta below the scale $\Lambda$. For
this reason, to describe nucleon dynamics one should use Eq.
(\ref{T2}) with the function $\Psi_0({\bf p})$ satisfying the two
conditions: it must have the low momentum expansion (\ref{Psi_0})
where second term is determined by the LEC $c_2$, and is related
to the LEC ${\cal J}_2$ by Eq. (\ref{J_2j}). These LEC's should be
derived from low energy experiment, and first of all from the $NN$
scattering data.

The scattering amplitude that results from the NLO $T$-matrix
shown in Eq. (\ref{T^2}) is
\begin{widetext}
\begin{eqnarray}\label{A^2}
A^{(2)}(p)&=&-\langle{\bf p}| T^{(2)}(z=E_p+i0)|{\bf p}\rangle
=-\frac{(1+c_2^*p^2)(1+c_2p^2)}{C^{-1}_0-M_0(z)(1+2\textsf{Re}c_2zm)+{\cal
J}_2zm}\Big(1+O\{Q^3/\Lambda^3\}\Big)\nonumber\\
&=&-\frac{4\pi}{m}\left[\frac{1}{\frac{4\pi}{m}C^{-1}_0+ip}+\frac{4\pi}{m}\frac{2C^{-1}_0
\textsf{Re} c_2-{\cal J}_2}{\{\frac{4\pi}{m}C^{-1}_0+ip\}^2}\cdot
p^2\right]\Big(1+O\{Q^3/\Lambda^3\}\Big)
\end{eqnarray}
\end{widetext}
Taking into account that for $p<<\Lambda$ this equation must
reproduce the effective range expansion
\begin{eqnarray}\label{ERE}
A(p)=-\frac{4\pi a}{m}[1-iap+(ar_0/2-a^2)p^2+O\{p^3\}],\nonumber
\end{eqnarray}
we get the relations between our LEC's and scattering length $a$
and the effective range $r_0$
\begin{eqnarray}\label{par_atNLO}
C_{0}=\frac{4\pi a}{m},
\quad 2\textsf{Re}c_2-\frac{4\pi a}{m}{\cal J}_2=\frac{ar_0}{2}.
\end{eqnarray}
Thus, at NLO we have two equations that relate the three LEC's
$C_0$, $c_2$ and ${\cal J}_2$ to the scattering data. This means
that the two nucleon scattering data are not sufficient to fix
these LEC's, and fitting to experimental data where the off-shell
properties of the $2N$ $T$-matrix manifest themselves is needed.
For  example, one can use the $pp$ bremsstrahlung observables that
are very sensitive on the low energy $NN$ interaction.

As we noted, for constructing the $T$-matrix at higher orders we
have to deal with the amplitude $T^{(2)}(z,{\bf p}_2,{\bf p}_1)$
given by Eq. (\ref{T^2(z,2,1)}). This amplitude contains the
functions $\Phi_0({\bf p}_2,{\bf p}_1)$, $\varphi_2({\bf p})$ and
$\Psi_2({\bf p})$ that  do not manifest themselves at NLO.These
functions should come into play at higher orders. At
next-to-next-to-leading order (NNLO) Eq. (\ref{dif_again}) takes
the form
\begin{eqnarray}
\frac{d T^{(4)}(z,{\bf p}_2,{\bf p}_1)}{dz} =
\Big[\tilde{\cal T}_0^{(4)}(z,{\bf p}_2){\cal T}_0^{(4)}(z,{\bf p}_1)\frac{dM_0(z)}{dz}\nonumber\\
+{\cal F}^{(2)}(z,{\bf p}_2,{\bf
p}_1)\Big]\left(1+O\left\{Q^5/\Lambda^5\right\}\right).\nonumber
\end{eqnarray}
Correspondingly, by using Eqs.  (\ref{calt2}),  (\ref{tilT2}), and
(\ref{T2}) for $t_{00}^{(4)}(z)$ we get  the equation
\begin{widetext}
\begin{eqnarray}
\frac{dt_{00}^{(4)}(z)}{dz} &=&
\Big[\left(t_{00}^{(4)}(z)\right)^2\frac{dM_0(z)}{dz}+\left(t_{00}^{(2)}(z)\right)^2\left\{\frac{dM_2(z)}{dz}
+\frac{dM_4(z)}{dz}-\int\frac{d^3k}{(2\pi)^3}\frac{\Psi_0^*({\bf
k}){\cal M}_2(z,{\bf k}) +\Psi_0({\bf k})\tilde {\cal M}_2(z,{\bf
k})}{(z-E_k)^2}\right\}\Big]\nonumber\\
&\times&\left(1+O\left\{Q^5/\Lambda^5\right\}\right) =
\Big[\left(t_{00}^{(4)}(z)\right)^2\frac{dM_0(z)}{dz}+\left(t_{00}^{(2)}(z)\right)^2\frac{d}{dz}\Big\{M_2(z)+
M_4(z)+{\cal C}_{2,2}(zm)^2M_0^2(z)\nonumber\\
&-&2\textsf{Re}{\cal A}_4(zm)^2M_0(z)+\textsf{Re}{\cal
B}_4(zm)^2\Big\}\Big]\Big(1+O\left\{Q^5/\Lambda^5\right\}\Big),\nonumber
\end{eqnarray}
where
\begin{eqnarray}\label{A&B}
{\cal A}_4 = m\int\frac{ d^3k}{(2\pi)^3}\frac{\Psi_0({\bf
k})\varphi_2^*({\bf k})}{k^{4}},\quad {\cal B}_4 = m\int\frac{
d^3k}{(2\pi)^3}\frac{\Psi_0^*({\bf k})\alpha_2({\bf k})}{k^{4}}.
\end{eqnarray}
 Solving this equation with the boundary condition
(\ref{C-0}) yields
\begin{eqnarray}
t_{00}^{(4)}(z)
=\frac{\left(1+O\left\{Q^5/\Lambda^5\right\}\right)}{C_0^{-1}-M_0(z)\Big\{1+2\textsf{Re}c_2zm+
\Big[|c_2|^2+2\textsf{Re}(c_4-{\cal A}_4)+{\cal
C}_{2,2}M_0(z)\Big](zm)^2\Big\}+{\cal J}_2zm+\Big[{\cal
J}_4-\textsf{Re}{\cal B}_4\Big](zm)^2}.\nonumber
\end{eqnarray}
This expression for $t_{00}^{(4)}(z)$ can than be used in the
equation for ${\cal T}_0^{(4)}(z,{\bf p})$
\begin{eqnarray}
\frac{d{\cal T}_0^{(4)}(z,{\bf p})}{dz} =
\Big[t_{00}^{(4)}(z){\cal T}_0^{(4)}(z,{\bf p})\frac{dM_0(z)}{dz}
+t_{00}^{(2)}(z){\cal T}_0^{(2)}(z,{\bf
p})\Big\{\frac{dM_2(z)}{dz}
+\frac{dM_4(z)}{dz}-\int\frac{d^3k}{(2\pi)^3}\frac{\Psi_0^*({\bf
k}){\cal M}_2(z,{\bf k})}{(z-E_k)^2}\Big\}\nonumber\\
-\Big[t_{00}^{(2)}(z)\Big]^2 \Psi_0({\bf
p})\int\frac{d^3k}{(2\pi)^3}\frac{\Psi_0({\bf k})\tilde{\cal
M}_2(z,{\bf k})}{(z-E_k)^2}
-t_{00}^{(2)}(z)\int\frac{d^3k}{(2\pi)^3}\frac{[\Psi_0({\bf
k})+{\cal M}_2(z,{\bf k})]\Phi_0({\bf k},{\bf
p})}{(z-E_k)^2}\Big]\Big(1+O\left\{Q^5/\Lambda^5\right\}\Big).
\nonumber
\end{eqnarray}
The solution of this equation with the boundary condition
(\ref{T(0,p)}) is
\begin{eqnarray}
{\cal T}_0^{(4)}(z,{\bf p})=t_{00}^{(4)}(z)\Big(\Psi_0({\bf
p})+{\cal M}_2(z,{\bf p})+{\cal M}_4(z,{\bf
p})-\frac{2}{5}(zm)^2M_0(z)\theta_4({\bf
p})+\frac{1}{2}(zm)^2\gamma_4({\bf p})\Big)
\left(1+O\left\{Q^5/\Lambda^5\right\}\right),\nonumber
\end{eqnarray}
where
\begin{eqnarray}
\theta_4({\bf p}) = m\int\frac{
d^3k}{(2\pi)^3}\frac{\varphi_2({\bf k})\Phi_0({\bf k},{\bf
p})}{k^{4}},\quad \gamma_4({\bf p}) = m\int\frac{
d^3k}{(2\pi)^3}\frac{\alpha_2({\bf k})\Phi_0({\bf k},{\bf
p})}{k^{4}}.\nonumber
\end{eqnarray}
In the same way for $\tilde{\cal T}_0^{(4)}(z,{\bf p})$ we get
\begin{eqnarray}
\tilde{\cal T}_0^{(4)}(z,{\bf
p})=t_{00}^{(4)}(z)\Big(\Psi^*_0({\bf p})+\tilde{\cal M}_2(z,{\bf
p})+\tilde{\cal M}_4(z,{\bf
p})-\frac{2}{5}(zm)^2M_0(z)\theta^*_4({\bf
p})+\frac{1}{2}(zm)^2\gamma^*_4({\bf p})\Big)
\left(1+O\left\{Q^5/\Lambda^5\right\}\right).\nonumber
\end{eqnarray}
Now we can rewrite the equation for $T^{(4)}(z,{\bf p}_2,{\bf
p}_1)$ in the form
\begin{eqnarray}
\frac{dT^{(4)}(z,{\bf p}_2,{\bf p}_1)}{dz} = \Big[\tilde{\cal
T}_0^{(4)}(z,{\bf p}_2){\cal T}_0^{(4)}(z,{\bf
p}_1)\frac{dM_0(z)}{dz}+\tilde{\cal T}_0^{(2)}(z,{\bf p}_2){\cal
T}_0^{(2)}(z,{\bf p}_1)\Big(\frac{dM_2(z)}{dz}
+\frac{dM_4(z)}{dz}\nonumber\\-\int\frac{d^3k}{(2\pi)^3}\frac{\Psi_0^*({\bf
k}){\cal M}_2(z,{\bf k})+\Psi_0({\bf k})\tilde {\cal M}_2(z,{\bf
k})}{(z-E_k)^2}\Big)-\tilde{\cal T}_0^{(2)}(z,{\bf p}_2)
\int\frac{d^3k}{(2\pi)^3}\frac{\Big(\Psi_0({\bf k})+{\cal
M}_2(z,{\bf k})\Big)\Phi_0({\bf k},{\bf
p}_1)}{(z-E_k)^2}\nonumber\\
-{\cal T}_0^{(2)}(z,{\bf p}_1)
\Big\{\int\frac{d^3k}{(2\pi)^3}\frac{\Phi_0({\bf p}_2,{\bf
k})\Big(\Psi_0^*({\bf k})+\tilde{\cal M}_2(z,{\bf
k})\Big)}{(z-E_k)^2}+\frac{d}{dz}zm\int\frac{d^3k}{(2\pi)^3}\frac{\Phi_0({\bf
p}_2,{\bf k})\Phi_0({\bf k},{\bf
p}_1)}{(z-E_k)k^2}\Big]\Big(1+O\left\{Q^5/\Lambda^5\right\}\Big).\nonumber
\end{eqnarray}
Solving this equation with the boundary condition
(\ref{boundcond}) yields
\begin{eqnarray}\label{T^4()}
T^{(4)}(z,{\bf p}_2,{\bf p}_1)&=&\tilde{\cal T}_0^{(4)}(z,{\bf
p}_2){\cal T}_0^{(4)}(z,{\bf
p}_1)\Big[t_{00}^{(4)}(z)\Big]^{-1}+\Phi_0({\bf p}_2,{\bf
p}_1)-zm\Upsilon_4({\bf p}_2,{\bf p}_1),
\end{eqnarray}
where $\Upsilon_4({\bf p}_2,{\bf
p}_1)=m\int\frac{d^3k}{(2\pi)^3}\frac{\Phi_0({\bf p}_2,{\bf
k})\Phi_0({\bf k},{\bf p}_1)}{k^4}$.
 In describing low energy nucleon dynamics the terms
including the function ${\cal M}_2(z,{\bf p})$ can be neglected,
and hence for $\langle {\bf p}_2|T^{(4)}(z)|{\bf p}_1\rangle$ we
can write
\begin{eqnarray}\label{expr_at N^2LO}
\langle {\bf p}_2|T^{(4)}(z)|{\bf
p}_1\rangle&=&\Big[\Big\{\Psi_0^*({\bf p}_2)+\tilde {\cal
M}_2(z,{\bf p}_2)\Big\}\Big\{\Psi_0({\bf p}_1)+{\cal M}_2(z,{\bf
p}_1)\Big\}t_{00}^{(4)}(z)+\Phi_0({\bf p}_2 ,{\bf p}_1
)\Big](1+O\{Q^5/\Lambda^5\}).
\end{eqnarray}
For momenta below $\Lambda$ this equation can be rewritten in the
form
\begin{eqnarray}\label{T^4}
\langle {\bf p}_2|T^{(4)}(z)|{\bf
p}_1\rangle%
&=&\Big[\Big(1+c_2^*p_2^2+c_4^*p_2^4+{\cal C}_{2,2}^*zmM_0(z)p_2^2
-{\cal A}_4^*zmp_2^2\Big)\Big(1+c_2p_1^2+c_4p_1^4+{\cal
C}_{2,2}zmM_0(z)p_1^2 -{\cal A}_4zmp_1^2\Big)\nonumber\\
&\times& t_{00}^{(4)}(z)+{\cal
C}_{2,2}p_2^2p_1^2\Big]\Big(1+O\{Q^5/\Lambda^5\}\Big).
\end{eqnarray}
\end{widetext}
Here we have used the fact that for $|{\bf p}|<\Lambda$ the
function $\alpha_2({\bf p})$ can be expanded as
$$\alpha_2({\bf p})={\cal A}_4p^2+{\cal A}_6p^4+...,\quad |{\bf p}|<\Lambda.$$
By using this equation and Eq. (\ref{expr_at N^2LO}), for the NNLO
$2N$ scattering amplitude we can write
\begin{eqnarray}
A^{(4)}(p)&=&-\frac{4\pi}{m}\frac{1}{a^{-1}+ip}\Big[1+\frac{r_0/2}{a^{-1}+ip}p^2\nonumber\\
&+&\frac{(r_0/2)^2}{(a^{-1}+ip)^2}p^4+\frac{r_1/2}{a^{-1}+ip}p^4+\ldots\Big]\nonumber
\end{eqnarray}
where $r_1=\frac{8\pi}{m}\cdot X$ with
$X=C_0^{-1}\Big(|c_2|^2+2\textsf{Re}(c_4-{\cal
A}_4)-(2\textsf{Re}c_2)^2+C_0^{-1}{\cal C}_{2,2}\Big)-{\cal
J}_4+\textsf{Re}{\cal B}_4+2{\cal J}_2\textsf{Re}c_2$. This is the
Kaplan-Savage-Wise (KSW) expansion of the effective theory
\cite{Kaplan2}. At the same time, Eq. (\ref{T^4}) allows one to
obtain the more general expansion
\begin{eqnarray}\label{A^4}
A(p)=-\frac{C_0(\mu)}{1+\frac{C_0(\mu)m}{4\pi}(\mu+ip)}-\frac{C_2(\mu)p^2}{[1+
\frac{C_0(\mu)m}{4\pi}(\mu+ip)]^2}\nonumber\\
+\frac{(C_2(\mu)p^2)^2m(\mu+ip)/4\pi}{[1+
\frac{C_0(\mu)m}{4\pi}(\mu+ip)]^3}-\frac{C_4(\mu)p^4}{[1+
\frac{C_0(\mu)m}{4\pi}(\mu+ip)]^2}+\ldots
\end{eqnarray}
with $C_0(\mu)=\Big[C_0^{-1}-\frac{m}{4\pi}\mu\Big]^{-1}$,
$C_2(\mu)=\Big(2C_0^{-1}\textsf{Re}c_2-{\cal J}_2\Big)C_0^2(\mu)$
and $C_4(\mu)=X\cdot C_0^2(\mu)+\frac{m}{4\pi}C_0\mu
C_2^2(\mu)/C_0(\mu)$, where $\mu$ is an arbitrary parameter.

We have demonstrated that, by using Eq. (\ref{dif_again}) and
taking into account that the $2N$ $T$-matrix in the ${}^1S_0$
channel should be of the form (\ref{anW}), one can construct this
$T$-matrix order by order in the expansion in powers of
$Q/\Lambda$. For example, the $2N$ $T$-matrix up to NLO and NNLO
is given by Eqs. (\ref{T2}) and (\ref{expr_at N^2LO}),
respectively. It should be noted that the way in which the above
expansion of the $2N$ $T$-matrix has been obtained is not a
regularization procedure. Moreover, the key point of our approach
is that the $2N$ $T$-matrix must necessarily satisfy the GDE
without regularization. This requirement directly follows from the
fact that at low energies nucleons emerge as the only effective
degrees of freedom, i.e., the probability to find a system, the
initially was in a two-nucleon state, in a state including other
degrees of freedom is negligible. Indeed, the above means that the
evolution operator defined on the $2N$ subspace of the Hilbert
space should be unitary. From this it follows immediately that the
$2N$ $T$-matrix must satisfy the GDE. This is because this
equation is an inevitable consequence of the requirement that the
evolution operator is of the form (\ref{ev}) and satisfies the
unitary condition. Our expansion of the $2N$ $T$-matrix is based
on the fact that the GDE must be satisfied at any order of the
expansion in powers of $Q/\Lambda$. Equation (\ref{dif_again})
represents the approximation to the GDE that allows one to obtain
the $2N$ $T$-matrix with an accuracy of order
$O\{(Q/\Lambda)^{2{\cal N}+1}\}$, if it is known with an accuracy
of order $O\{(Q/\Lambda)^{2{\cal N}-1}\}$.  As we have shown, in
this way we get the same expansion of the scattering amplitude
that in the standard pionless EFT is derived by summing the bubble
diagrams and performing regularization and renormalization. The
advantage of the above way of the expansion of the effective
theory of nuclear forces is that it allows one to construct not
only the scattering amplitude but also the off-shell $T$-matrix.
As we show in the next section, this provides a new way to
formulate the effective theory of nuclear forces.

\section{The effective theory of nuclear forces.}
As follows from Eq. (\ref{evo}), in order to describe the low
energy dynamics we have to obtain the $T$-matrix for any $z$
relevant for the low energy theory. Equation (\ref{difer}) allows
one to obtain the operator $T(z)$ for any $z$ provided that some
boundary condition on this equation is specified. The boundary
condition (\ref{T(z)to}) means that the most of the contribution
to the operator $T(z)$ in the limit $|z|\to\infty$ comes from the
processes associated with the fundamental interaction in the
system under study  that are described by the operator $B(z)$. On
the other hand, for the low energy theory to be consistent, the
operator $B(z)$ must be determined in terms of low energy degrees
of freedom. This means that, despite the boundary condition
(\ref{T(z)to}) formally implies that $z$ must be let to infinity,
really one has to restrict oneself to a "high" energy region which
 is much above the low energy
scale but is well below the scale of the underlying high energy
physics. We will use ${\cal D}$ to denote this "high" energy
region. The fact that the energy region ${\cal D}$ lies much above
the scale of the low energy dynamics  implies that at such
"infinite" energies the main part of the contribution to the
operator $T(z)$ comes from processes that can be thought of as a
"fundamental" interaction in the low energy theory, and, as a
result, the interaction operator $B(z)$ is  close enough to the
true $2N$ $T$-matrix. Thus, in order to take into account the fact
that any theory has a range of validity, instead of the condition
(\ref{T(z)to}), we have to use the following boundary condition on
Eq. (\ref{difer}):
\begin{eqnarray}
 \langle \psi_2|T(z)|\psi_1\rangle &=& \langle \psi_2|B(z)|\psi_1\rangle\label{T(z=s)}
+O\left\{h(z)\right\},  z\in{\cal D}.
\end{eqnarray}
In order for Eq. (\ref{difer}) with the boundary condition
(\ref{T(z=s)}) to have a unique solution, the interaction operator
must be close enough inside domain ${\cal D}$ to the true $2N$
$T$-matrix, i.e., the function $h(z)$ must be sufficiently small
for $z\in{\cal D}$. This means that the operator $B(z)$ must
satisfy the equation
\begin{eqnarray}
\frac{d \langle \psi_2|B(z)|\psi_1\rangle}{dz} &=& - \sum
\limits_{n} \frac{\langle \psi_2|B(z)|n\rangle\langle n |B(z)|\psi_1\rangle}{(z-E_n)^2}\nonumber\\
 &+&O\left\{|z|^{-1}h(z)\right\},\quad z\in{\cal D}. \label{diferB2}
\end{eqnarray}

 In the low energy theory of nuclear forces all
processes that in ChPT are described by irreducible diagrams
involving only two external nucleons can be considered as such
"fundamental" processes. Here irreducible diagrams are $2N$
irreducible: Any intermediate state contains at least one pion or
isobar. It is natural to expect that the relative contribution of
reducible $2N$ diagrams tends to zero as $z$ increases, and, in a
region that lies much above the scale of low energy nucleon
dynamics, the main part of the contribution comes from the
"fundamental" processes that are described by the $2N$ irreducible
diagrams, and therefore the interaction operator $B(z)$ should
describe the contributions from these processes.  On the other
hand, because of the separation of scales provided by QCD the
above "high" energy region of the low energy theory lies still
much below the scale of the underlying physics. In other words,
the amplitudes $\langle \psi_2|B(z)|\psi_1\rangle$ that constitute
the interaction operator and generate low energy nucleon dynamics
really are low energy (in the scale of the underlying theory)
amplitudes that in ChPT are described by the irreducible $2N$
diagrams. In principle they can be obtained within the underlying
high energy theory in some low energy limit, and can be directly
used as  building blocks for constructing the low energy theory.
Thus the GQD allows one to build a bridge between QCD and low
energy nucleon dynamics. It is hoped that in the future it will be
possible to obtain the amplitudes $\langle
\psi_2|B(z)|\psi_1\rangle$ in terms of QCD with such accuracy that
the corresponding operator $B(z)$ will determine a unique low
energy $2N$ $T$-matrix.  There is no reason to believe that this
operator must necessarily generate the Hamiltonian low energy
dynamics. In fact, as it follows from Eq. (\ref{B(z)-H_I}), for
this $\langle {\bf p}_2|B(z)|{\bf p}_1\rangle$ must have a
negligible dependence on $z$ inside  the domain ${\cal D}$.
However, as we show below, such a behavior of $\langle {\bf
p}_2|B(z)|{\bf p}_1\rangle$ is at variance with the symmetries of
QCD. On the other hand, in the region ${\cal D}$ QCD must
reproduce the  $2N$ $T$-matrix that satisfies the GDE and hence
can be used as a  $NN$ interaction operator governing low energy
dynamics.

In the case of the pionless theory the problem of constructing the
generalized interaction operator actually is  a special case of
the problem of obtaining the $2N$ $T$-matrix when we restrict
ourselves to the domain ${\cal D}$. We  should also keep in mind
that, as it follows from Eq. (\ref{T(z=s)}), in the domain ${\cal
D}$ the operator $B(z)$ coincides with the true $T$-matrix with an
accuracy of order $O\{h(z)\}$. In other words, the operator
$B_{eff}^{(2)}(z)$ that describes the $NN$ interaction up to
next-to-leading order is related to the operator $T^{(2)}(z)$ in
the domain ${\cal D}$ by
\begin{widetext}
\begin{eqnarray}
\langle {\bf p}_2|T^{(2)}(z)|{\bf p}_1\rangle=\frac{\Psi_0^*({\bf
p_2} )\Psi_0({\bf p_1}
)}{C_0^{-1}-M_0(z)-M_2(z)}\Big(1+O\{Q^3/\Lambda^3\}\Big)=-\Psi_0^*({\bf
p_2} )\Psi_0({\bf
p_1})\nonumber\\
\times\Big[M_0^{-1}(z)\Big(1-2\textsf{Re}c_2zm\Big)+M_0^{-2}(z)\Big(C_0^{-1}+{\cal
J}_2zm\Big) \Big]
\left(1+O\{M_0^{-3}(z)C_0^{-2}\}\right)\left(1+O\{Q^3/\Lambda^3\}\right)\nonumber\\=\langle
{\bf p}_2|B_{eff}^{(2)}(z)|{\bf
p}_1\rangle\left(1+O\{M_0^{-3}(z)C_0^{-2}\}\right),\quad z\in{\cal
D},\nonumber
\end{eqnarray}
\end{widetext}
with
\begin{eqnarray}\label{B_at NLO}
\langle {\bf p}_2|B_{eff}^{(2)}(z)|{\bf p}_1\rangle=-\Psi_0^*({\bf
p_2} )\Psi_0({\bf p_1}
)\nonumber\\
\times\Big[M_0^{-1}(z)\Big(1-2\textsf{Re}c_2zm\Big)+M_0^{-2}(z)\Big(C_0^{-1}+{\cal
J}_2zm\Big) \Big] \nonumber\\
\times\left(1+O\{Q^3/\Lambda^3\}\right).
\end{eqnarray}
 In this equation the effects of the high energy physics are parametrized by
the LEC's $C_0$, $c_2$, and ${\cal J}_2$. These LEC's put the
constraints on the form factor $\Psi_0({\bf p})$:  The LEC $c_2$
determines its low momentum properties of the form factor
$\Psi_0({\bf p} )$ with an accuracy of order $(Q/\Lambda)^3$,
while the constant ${\cal J}_2$ determines  integral properties of
this function relevant for calculations at the same order. The
above means that, in contrast with the operator $B(z)$ which
unequally determines the relevant solution of Eq. (\ref{difer}),
the effective interaction operator $B_{eff}^{(2)}(z)$ determines
the set $\Omega_2$ of the solutions  of this equation that
coincide with the true $2N$ $T$-matrix with an accuracy of order
$O\{(Q/\Lambda)^3\}$. From the above expansion of $T^{(2)}(z,{\bf
p}_2,{\bf p}_1)$ it follows that the function $h(z)$ which
determines the size of the terms in the interaction operator that
can be omitted should be of order
\begin{equation}
  h(z)=O\{|C_0^{-2}M_0^{-3}(z)|\}.\nonumber
\end{equation}
The effective interaction operator $B_{eff}^{(2)}(z)$ given by Eq.
(\ref{B_at NLO}) actually represents a set of the interaction
operators that generate the $T$-matrices belonging to the solution
set $\Omega_2$. Each of these operators corresponds to some
definite function $\Psi_0({\bf p} )$ satisfying the above
requirement. Such an operator is of the form
\begin{eqnarray}\label{B(z)}
\langle{\bf p_2}|B(z)|{\bf p_1}\rangle=-\Psi_0^*({\bf p_2}
)\Psi_0({\bf p_1} )[M_0^{-1}(z)\nonumber\\+M_0^{-2}(z)C_0^{-1}],
\end{eqnarray}
and generates the $T$-matrix
\begin{eqnarray}\label{T(z)}
\langle{\bf p_2}|T(z)|{\bf p_1}\rangle=\frac{\Psi_0^*({\bf p_2}
)\Psi_0({\bf p_1} )}{C_0^{-1}-M_0(z)-M(z)}
\end{eqnarray}
where $M(z)$ is given by Eq. (\ref{Mnz}). It should be noted that
in contrast with the effective interaction $B_{eff}^{(2)}(z)$ the
operator (\ref{B(z)}) contains only two terms $M_0^{-1}(z)$ and
$M_0^{-2}(z)C_0^{-1}$ that determine its dependence on $z$. This
is because the form factor $\Psi_0({\bf p})$ that occurs in Eq.
(\ref{B(z)}) is assumed to be known, and the operator (\ref{B(z)})
is close enough in the domain ${\cal D}$ to some separable
$T$-matrix with this form factor to unequally determine it. The
function $M(z)$ describing the dependence of this $T$-matrix on
$z$ is completely determined by the form factor $\Psi_0({\bf p})$.
From the fact that the form factor satisfies the above
requirements and hence the $T$-matrix (\ref{T(z)}) belongs to the
solution set $\Omega_2$ corresponding to given LEC's $c_2$
  and ${\cal J}_2$, it follows that the first term in the expansion of
  $M(z)$ shown in Eq. (\ref{Mnz})
\begin{eqnarray}\label{exp_M(z)}
  M(z)=\left[2\textsf{Re}c_2zmM_0(z)-{\cal
  J}_2zm\right]
  \left(1+O\{Q^3/\Lambda^3\}\right)\nonumber
\end{eqnarray}
corresponds to the same constants. In contrast, the effective
interaction operator $B_{eff}^{(2)}(z)$ determines the solution
set $\Omega_2$ as a whole not the solution corresponding to a
given form factor, and the dependence of this operator on $z$ that
is described by the term $zm{\cal
  J}_2-2\textsf{Re}c_2zmM_0(z)$ puts the above constraints on the
  function $\Psi_0({\bf p} )$.

Equation (\ref{expr_at N^2LO}) represents the set $\Omega_4$ of
the solutions of the GDE that coincide with the true $2N$
$T$-matrix with an accuracy of order $O\{Q^5/\Lambda^5\}$. This
solution set is determined by the values of the LEC's $C_0$,
$c_2$, $c_4$, ${\cal C}_{2,2}$, ${\cal J}_2$, ${\cal J}_4$, ${\cal
A}_4$ and ${\cal B}_4$. These LEC's put the constraints on the
functions $\Psi_0({\bf p})$ and $\Phi_0({\bf p}_2,{\bf p}_1)$ that
determine the definite solutions belonging to the set $\Omega_4$:
The constants $c_2$, $c_4$ and ${\cal C}_{2,2}$ that occur in
their low momentum expansions shown in Eqs. (\ref{Psi_0}) and
(\ref{Phi}) have the same values as the above LEC's, and these
functions have the integral properties that are parametrized via
Eqs. (\ref{J_2j}) and (\ref{A&B}) by the LEC's ${\cal J}_2$,
${\cal J}_4$, ${\cal A}_4$ and ${\cal B}_4$. The effective
interaction operator that generates the solution set $\Omega_4$ of
the GDE is of the form
\begin{widetext}
\begin{eqnarray}\label{B^4(z)}
\langle{\bf p_2}|B_{eff}^{(4)}(z)|{\bf
p_1}\rangle=\Big[-\{\Psi_0^*({\bf p_2})+\tilde{\cal M}_2(z,{\bf
p}_2)\}\{\Psi_0({\bf p_1})+{\cal M}_2(z,{\bf p}_1)\}\nonumber\\
\times \Big\{M_0^{-1}(z)\Big(1-2\textsf{Re}c_2zm-\Big[
|c_2|^2+2\textsf{Re}(c_4-{\cal A}_4)-(2\textsf{Re}c_2)^2+{\cal
C}_{2,2}M_0(z)\Big](zm)^2\Big)\nonumber\\+M_0^{-2}(z)\Big(C_0^{-1}+{\cal
J}_2zm+\Big[{\cal J}_4-\textsf{Re}{\cal
B}_4\Big](zm)^2\Big)\Big\}+\Phi_0({\bf p}_2,{\bf p}_1)\Big]
\Big(1+O\{Q^5/\Lambda^5\}\Big).
\end{eqnarray}
\end{widetext}
This equation represents the set of the interaction operators that
generate the solutions of the GDE belonging to the set $\Omega_4$,
and for calculations with an accuracy of order
$O\{Q^5/\Lambda^5\}$ one can use the operator (\ref{B^4(z)}) with
any functions $\psi_0(p )$ and $\Phi_0({\bf p}_2 ,{\bf p}_1 )$
satisfying the above conditions with the corresponding LEC's. In
the same way we can obtain the set $\Omega_{2{\cal N}}$ of the
solutions that coincides with the true $2N$ $T$-matrix with an
accuracy of order $O\{(Q/\Lambda)^{2{\cal N}+1}\}$ for any ${\cal
N}$, and as ${\cal N}$ increases one approaches to the true $2N$
$T$-matrix closer and closer.

As we have noted the formalism of the GQD prescribes to start with
the analysis of the theory at high energies where in general the
structure of the theory is most simple, but in the pionless theory
this is not the case. In this theory the Weinberg analysis of the
diagrams in ChPT yields the same expansion (\ref{anW}) of the $2N$
$T$-matrix for all relevant energies, and one can use it for
obtaining this $T$-matrix not only in the domain ${\cal D}$ but
also at low energies. However, in the case when pions must be
included as explicit degrees of freedom the structure of the $2N$
$T$-matrix can be derived directly from the analysis of the
diagrams in ChPT only at "high" energies belonging to the domain
${\cal D}$, where the theory has the most simple structure.
Indeed, in this case the $T$-matrix given by Eq. (\ref{T(z=s)})
should be considered as the contribution from all relevant contact
diagrams in ChPT, and represents not the full $2N$ $T$-matrix but
only its short-range component. The long-range component should be
obtained by iterating the pion exchange potential that
parametrizes the long-range part of the $NN$ interaction. However,
only at high energies the full $2N$ $T$-matrix can be represented
as a sum of these contributions. This point will be demonstrated
in the next section.

\section{A new look at the Weinberg program}

The Weinberg proposal was based on the assumption that the only
equation that can govern low energy nucleon dynamics is the LS
(the Schr{\"o}dinger) equation, and hence what one has to derive
from the analysis of diagrams in ChPT is an $NN$ potential.
However, there is no reason to consider that low energy nucleon
dynamics is necessarily governed by the Schr{\"o}dinger equation.
In principle this dynamics may be governed by the GDE with a
nonlocal-in-time interaction operator when this equation is not
equivalent to the Schr{\"o}dinger equation. In fact, as has been
shown in Ref. \cite{R.Kh.:1999}, only the GDE must be satisfied in
any case, not the Schr{\"o}dinger equation. In the light of this
fact the Weinberg program can be considered from a new point of
view: Instead of the Schr{\"o}dinger (LS) equation, one should use
the GDE. Below we show that the approach based on this more
general equation allows a modification of the chiral potential
model that can increase its predictive power.

   The leading order
$NN$ potential that has been derived from the Weinberg analysis of
diagrams in ChPT is \cite{EFT3}
\begin{eqnarray}\label{V_}
V({\bf p}_2,{\bf p}_1)=V_\pi({\bf p}_2,{\bf
p}_1)+C_S+C_T{\bf{\sigma_1}},\nonumber\\
V_\pi({\bf p}_2,{\bf
p}_1)=-\left(\frac{g_A^2}{2f_\pi^2}\right)\frac{\bf{q}
\cdot{\bf{\sigma_1}}{\bf{q}} \cdot{\bf \sigma}_2}
{q^2+m_\pi^2}{\bf\tau_1}\cdot{\bf{\tau}_2},
\end{eqnarray}
with ${\bf q}\equiv{\bf p}_2-{\bf p}_1$. The coupling $g_A$ is the
axial coupling constant, $m_\pi$ is the pion mass, $f_\pi$ is the
pion decay constant, and ${\bf \sigma}({\bf \tau})$ are the Pauli
matrices acting in spin (isospin) space. The derivation of the
chiral $NN$ potential at higher orders was pioneered by
Ord{\'o}{\~n}ez, Ray and van Kolck \cite{Ordonez,van Kolck} who
derived a $NN $ potential in coordinate space based upon ChPT at
NNLO. Epelbaum et al. \cite{Epelbaum} constructed the first
momentum-space $NN$ potential at NNLO. Recently the first $NN$
potential at fourth order ($N^3LO$) of ChPT was developed by Entem
and Machleidt \cite{Entem2}. This potential was shown to reproduce
the $NN$ data below 290 MeV with the same accuracy as
phenomenological high-precision potentials. This potential can be
written in the form
\begin{eqnarray}\label{chipot}
V=V_\pi+V_{sh},\\
V_\pi=V_\pi^{(0)}+V_\pi^{(2)}+V_\pi^{(3)}+V_\pi^{(4)},
\end{eqnarray}
and
\begin{eqnarray}
V_{sh}=V_{sh}^{(0)}+V_{sh}^{(2)}+V_{sh}^{(4)}.
\end{eqnarray}
Here $V_\pi^{(\nu)}$ is the contributions to the long-distance
part $V_\pi$ of the chiral potential at order $\nu$, and
$V_{sh}^{(\nu)}$ is the $\nu$-th order contribution to the
short-distance part. At very low energies the pion-exchange part
of the $NN$ interaction may be included into the contact term, and
the $NN$ potential in the ${}^1S_0$ channel takes the form
\begin{equation}\label{expot}
\langle{\bf p}_2|V|{\bf
p}_1\rangle=c_0+c_2(p_2^2+p_1^2)+c_4(p_2^4+p_1^4)...
\end{equation}
At NLO one can restrict oneself to the first two terms in the
expansion (\ref{expot}), and the potential takes the form
\begin{equation}\label{NLO}
\langle{\bf p}_2|V|{\bf p}_1\rangle=c_0+c_2(p_2^2+p_1^2).
\end{equation}
The potential (\ref{NLO}) is very singular at short distances, and
the LS equation with this potential makes no sense without
regularization. Let us perform regularization by using a momentum
cutoff. In this case the singular potential (\ref{NLO}) is
replaced by the regularized one $\langle{\bf p}_2|V_{\Lambda}|{\bf
p}_1\rangle=f^*(p_2/\Lambda)\langle{\bf p}_2|V|{\bf p}_1\rangle
f(p_1/\Lambda)$, where the form factor $f(p )$ satisfies $f(0)=1$
and falls off rapidly for $p
>1$.  At next-to-leading order the regularized $NN$ potential in
the ${}^1S_0$ channel is of the form
\begin{equation}\label{regNLO}
\langle{\bf p}_2|V_\Lambda^{(2)}|{\bf
p}_1\rangle=f^*(p_2/\Lambda)\Big(c_0(\Lambda)+c_2(\Lambda)(p_2^2+p_1^2)\Big)f(p_1/\Lambda).
\end{equation}
This potential has a two-term separable form and so the
corresponding LS equation can be solved using standard techniques.
The $T$-matrix obtained in this way is
\begin{widetext}
\begin{eqnarray}
\langle{\bf p_2}|T_{\Lambda}(z)|{\bf
p_1}\rangle=\frac{c_0(\Lambda)+c_2^2(\Lambda)I_4(z)+\Big(c_2(\Lambda)-c_2^2(\Lambda)I_2(z)\Big)
\Big(p_1^2+p_2^2\Big)+c_2^2(\Lambda)I_0(z)p_1^2p_2^2
}{1-\Big(c_0(\Lambda)+c_2^2(\Lambda)I_4(z)\Big)I_0(z)-2c_2(\Lambda)I_2(z)+c_2^2(\Lambda)I_2^2(z)}
f^*(p_2/\Lambda)f(p_1/\Lambda).\nonumber
\end{eqnarray}
where the integrals $I_n(z)$ are given by
$$I_{2n}(z)=\int\frac{d^3k}{(2\pi)^3}\frac{k^{2n}}{(z-E_k)}\left|f(k/\Lambda)\right|^2.$$ This expression can
be rewritten in the form
\begin{eqnarray}
\langle{\bf p_2}|T_{\Lambda}(z)|{\bf
p_1}\rangle=\frac{1+\Big(\Big[c_2(\Lambda)-c_2^2(\Lambda)I_2(z)\Big]
\left(p_1^2+p_2^2\right)+c_2^2(\Lambda)I_0(z)p_1^2p_2^2\Big)\Big[c_0(\Lambda)+c_2^2(\Lambda)I_4(z)\Big]^{-1}
}{\Big(1-2c_2(\Lambda)I_2(z)+c_2^2(\Lambda)I_2^2(z)\Big)
\Big[c_0(\Lambda)+c_2^2(\Lambda)I_4(z)\Big]^{-1}-I_0(z)}
f^*(p_2/\Lambda)f(p_1/\Lambda).\nonumber
\end{eqnarray}
\end{widetext}
Taking into account that $I_n(z)=zmI_{n-1}(z)+I_n(0)$ and
\begin{equation}\label{I(0)}
 I_{2n}(0)=-\frac{m\Lambda^{2n+1}}{(2\pi)^3}\int d^3\tilde
 k\tilde{k}^{2(n-1)}|f(\tilde k)|^2,
\end{equation}
and keeping only terms that contribute at leading and
next-to-leading orders, we can write
\begin{eqnarray}\label{Tpol}
&&\langle{\bf p_2}|T_{\Lambda}(z)|{\bf
p_1}\rangle\nonumber\\&=&\frac{(1+c_2p_1^2)(1+c_2p_2^2)(1
+O\{(Q/\Lambda)^5\})}{C_0^{-1}
-\frac{m}{4\pi}\sqrt{-zm}\left(1+2c_2zm\right)+{\cal J}_2zm},
\end{eqnarray}
where
\begin{eqnarray}
C_0=\left[c_0(\Lambda)+c_2^2(\Lambda)I_4(0)\right]
[1-\left(c_0(\Lambda)+c_2^2(\Lambda)I_4(0)\right)\nonumber\\\times
I_0(0)-2c_2(\Lambda)I_2(0)+c_2^2(\Lambda)I_2^2(0)]^{-1},\nonumber
\end{eqnarray}
\begin{eqnarray}\label{c_2}
c_2=-\frac{c_2(\Lambda)\Big(1-c_2(\Lambda)I_2(0)\Big)}{c_0(\Lambda)+c_2^2(\Lambda)I_{4}(0)},
\end{eqnarray}
\begin{eqnarray}\label{J_1}
{\cal J}_2=-2c_2I_{0}(0)-c_2^2I_{2}(0).
\end{eqnarray}
Note that, since the parameters $c_0(\Lambda)$ and $c_2(\Lambda)$
that appear in Eq. (\ref{regNLO}) are assumed to be real, the
constant $c_2$ is also real.

 This is just the expression for the $T$-matrix
shown in Eq. (\ref{T^2}) with given LEC's. Thus, even if we start
from the LS equation with the singular potential (\ref{NLO}),
after regularization and renormalization we arrive at the
expression (\ref{T^2}) for the NLO $2N$ $T$-matrix we have derived
from the requirement that the $T$-matrix is of the form
(\ref{anW}) and satisfies the GDE. The constants that occur in the
potential (\ref{NLO}) should be fitted to the scattering length
and effective range. As it follows from Eq. (\ref{par_atNLO}),
this means that the constants $C_0$ is fixed by the scattering
length $C_0=4\pi a/m$, and $c_2$ and ${\cal J}_2$ must satisfy the
equation
\begin{eqnarray}\label{Rec_2}
{\cal J}_2=C_0^{-1}\left(2c_2-\frac{ar_0}{2}\right).
\end{eqnarray}
In this way, by using the regulator function of the form
\begin{equation}\label{regm}
f(p/\Lambda)=\exp(-p^4/\Lambda^4),
\end{equation}
with $\Lambda=150$ $MeV$, for $c_2$ and ${\cal J}_2$  we get the
following values of the LEC's: $c_2=-4.28\cdot
10^{-5}\text{MeV}^{-2}$, ${\cal J}_2=-0.463$. The half off-shell
$T$-matrix reproduced by the solution set $\Omega_2$ of the GDE
corresponding to these constants is shown in Figs. 1. As we see,
the NLO amplitude $A(p',p)=-\langle {\bf p'}|T(E_p+i0)|{\bf
p}\rangle$ is much larger than the leading order one even for very
low momentum ${\bf p'}$. This may be a signal that the EFT
expansion breaks down in describing the off-shell behavior of the
$2N$ $T$-matrix at anomal low energies. Our finding is consistent
with results obtained by Kaplan, Savage, and Wise \cite{Kaplan}:
The expansion of the $NN$ potential
\begin{figure}
\resizebox{1\columnwidth}{!}{\includegraphics{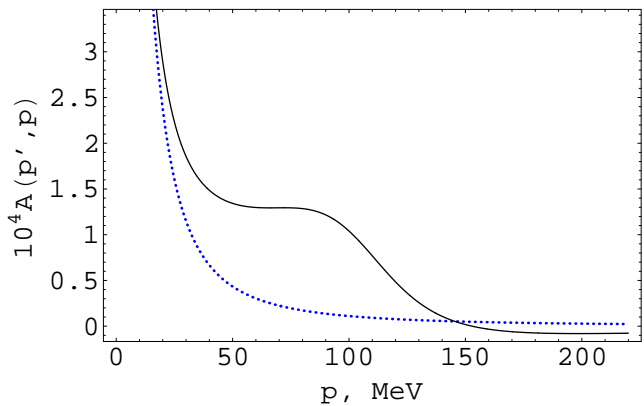}} \caption{The
real part of the half off-shell $T$-matrix as a function of the
on-shell momenta $p$. The curves represent the real part of the
amplitude $A(p',p)=-\langle {\bf p}'|T(E_{p}+i0)|{\bf p}\rangle$
($|{\bf p}'|=10$ $MeV$) obtained by solving the LS equation with
the potential (\ref{regNLO}) at LO (dotted line) and NLO (solid
line). The exponential regulator was employed with
$\Lambda=150{\text {MeV}}$.}
\end{figure}
\begin{figure}
\resizebox{1\columnwidth}{!}{\includegraphics{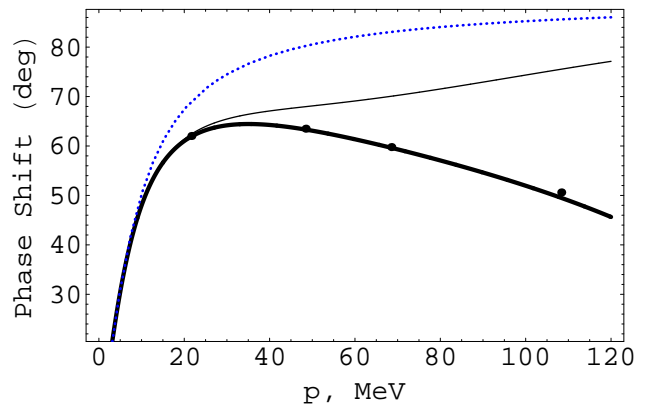}} \caption{The
${}^1S_0$ phase shift of $NN$ scattering. Solid dots represent the
experimental data \cite{Nijmegen}. The dotted, solid and heavy
solid curves represent the results obtained by using Eq.
(\ref{T^4(z)}) with the LEC's $c_2=8\cdot 10^{-5}\text{MeV}^{-2}$,
${\cal J}_2=-0.616$ and $c_4=-2\cdot 10^{-9} {\text
MeV}^{-4},\quad {\cal J}_4=-8\cdot 10^{-5} {\text MeV}^{-2}$ at
LO, NLO and NNLO, respectively.}
\end{figure}
\begin{figure}
\resizebox{1\columnwidth}{!}{\includegraphics{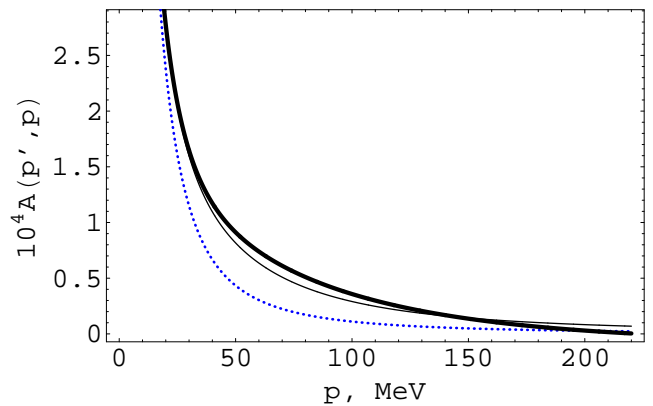}} \caption{The
real part of the half off-shell $T$-matrix as a function of the
on-shell momenta $p$. The dotted, solid and heavy solid curves
represent the solutions of the GDE for LEC's $c_2=8\cdot
10^{-5}\text{MeV}^{-2}$, ${\cal J}_2=-0.616$ and $c_4=-2\cdot
10^{-9} {\text MeV}^{-4},\quad {\cal J}_4=-8\cdot 10^{-5} {\text
MeV}^{-2}$ at LO, NLO and NNLO, respectively.}
\end{figure}
 $V=\sum\limits_{\mu=0}^\infty V_\mu$ with $V_\mu$
being contact potentials breaks down at the scale $\Lambda\sim
1/\sqrt{ar_0}$ since at this scale $V_0\simeq  V_2$. This result
is very discouraging from the EFT point of view: The amplitudes
should be expanded in powers of $(Q/\Lambda)^\nu$ where $\Lambda$
is a mass scale typical of the particles not included explicitly
in the theory. In the low energy EFT, with the pion integrated
out, one would expect $\Lambda\sim m_\pi$. The scale $\Lambda\sim
1/\sqrt{ar_0}$ at which the effective expansion of the $2N$
potential breaks down can hardly be called a typical QCD scale.
For this reason in Ref. \cite{Kaplan} it was proposed to expand
not the potential but the Feynman amplitudes. This expansion was
shown to converge up to the scale $\Lambda\sim m_\pi$. However, in
this way one can calculate only the scattering amplitudes not the
off-shell $T$-matrix. The advantage of the way of the expansion of
the effective theory based on the formalism of the GQD is that it
allows one to calculate not only the scattering amplitudes but
also the off-shell $2N$ $T$-matrix, and its EFT expansion
converges up to the expected scale set by the pion mass. But for
this we must not restrict ourselves to the solution sets
associated with potential model when the parameters $c_2$ and
${\cal J}_2$ are constrained not only by Eq. (\ref{regm}) but also
by Eq. (\ref{c_2}). Indeed, in general these LEC's determining the
solution set $\Omega_2$ must satisfy only Eq. ({\ref{regm}). This
means that fitting to the scattering data does not allow one to
obtain these constants. However, it is natural to expect that the
off-shell $T$-matrix must approach order by order to the true one
with the same accuracy as the on-shell matrix elements that
determine the phase shift. This puts some  additional constraint
on the values of the  constants. Let us investigate the problem of
the convergence of the effective expansion up to NNLO. At this
order the $2N$ $T$-matrix is given by Eq. (\ref{T^4}). For
simplicity we will restrict ourselves to the case when the
function $\Phi_0({\bf p}_2,{\bf p}_1)$ in Eq. (\ref{B^4(z)}) is
equal to zero, and the expression for the NNLO $2N$ $T$-matrix in
the ${}^1S_0$ channel takes the form
\begin{widetext}
\begin{eqnarray}\label{T^4(z)}
\langle{\bf p_2}|T^{(4)}(z)|{\bf
p_1}\rangle=\frac{(1+c_2^*p_2^2+c_4^*p_2^4)(1+c_2p_1^2+c_4p_1^4)}
{C_0^{-1}-\frac{m}{4\pi}\sqrt{-zm}\Big(1+2\textsf{Re}c_2zm+
\Big[|c_2|^2+2\textsf{Re}c_4\Big](zm)^2\Big)+{\cal J}_2zm+{\cal
J}_4(zm)^2}\nonumber\\
\times\left(1+O\{Q^5/\Lambda^5\}\right),\quad |{\bf
p}_1|<\Lambda,\quad |{\bf p}_2|<\Lambda.
\end{eqnarray}
\end{widetext}
This is a good approximation to the $2N$$T$-matrix because it has
a pole near $E=0$ in the ${}^1S_0$ channel. The results of our
calculations have shown that an acceptable convergence of the
effective expansion is achieved in the case when the parameter
$c_2$ is positive. For example, as we see from Figs. 3 and 4,
convergence up to the expected scale takes place for the following
set of the LEC's: $C_0= -0.0016\text{MeV}^{-2}$ (fixed at LO);
$c_2=8\cdot 10^{-5}\text{MeV}^{-2}$ and ${\cal J}_2=-0.616$ (fixed
at NLO); $c_4=-2\cdot 10^{-9} {\text MeV}^{-4}$ and  ${\cal
J}_4=-8\cdot 10^{-5} {\text MeV}^{-2}$} (fixed at NNLO). Note that
in this case we deal with the set of the solutions of the GDE that
cannot be associated with any potential. This is because the
condition (\ref{J_1}) that can be satisfied in the case of the
solutions obtained starting from the LS equation with the singular
potential (\ref{NLO}) allows only negative $c_2$. In fact, taking
into account that in the ${}^1S_0$ channel
$\frac{1}{2}ar_0=-(\frac{1}{35\text{MeV}})^2$, while $c_2$ is of
order $\frac{1}{\Lambda^2}\sim(\frac{1}{m_\pi})^2$, from Eq.
(\ref{Rec_2}) we conclude that ${\cal J}_2$ is close to
$-\frac{1}{2}ar_0C_0^{-1}$, and hence is negative. From Eqs.
(\ref{I(0)}) and (\ref{J_1}) than it follows that the constant
must be negative.

For momentum $p\sim m_\pi$ explicit pion fields must be included
in the theory. In this case in analyzing the time ordered diagrams
for $2N$ $T$-matrix one has to take into account that the
contributions from the "$2N$ irreducible" diagrams manifest
themselves not only as contact interactions but also as
pion-exchange ones. In this context the $2N$ $T$-matrix obtained
in the previous section should be considered as a contribution
from the processes in which only contact interactions come into
play. According to the Weinberg proposal this contribution  should
be obtained by iterating the corresponding contact potential  in
the LS equation. On the other hand, the solutions of the LS
equation have the following asymptotic behavior
$$\langle {\bf p}_2|T(z)|{\bf p}_1\rangle\tend\limits_{|z|\to\infty}\langle {\bf p}_2|V|{\bf p}_1\rangle,$$
which means that in the limit of high energies the $T$-matrix gets
the dominant contribution from the potential. In our case, when
this $T$-matrix describes the sum of contributions from the all
relevant contact diagrams in ChPT and the potential is its  "$2N$
irreducible"  part, the above should mean that this part gives the
main contribution in the limit of high energies. However, this is
not the case because all Feynman diagrams with $C_0$'s at each
vertex give a contribution of the same order. In the chiral
potential approach this problem is solved by putting in $ad$ $hoc$
short-ranged form factors in front of the contact interactions. As
we have demonstrated, in the ${}^1S_0$ channel at NLO after
renormalization of the solution of the LS equation with potentials
regularized in this way we arrive at the expression (\ref{T2}) we
have obtained, by using the GDE and the momentum dependence of the
above sum of the contact diagrams  that is given by Eq.
(\ref{anW}) without extracting any potential. It can also be shown
that renormalization of the solution of the LS equation with the
NNLO contact potential gives rise to the expression
(\ref{B^4(z)}). However, as we have seen, there are solutions of
the GDE consistent with symmetries of QCD that cannot be obtained
by starting from the LS equation with regularized contact
potentials. This is not surprising for the following reasons.

As is well known, the short-distance physics in ChPT is not the
full short-distance physics of QCD. At distance scales $\sim
1/\Lambda_\chi$ mechanisms not encoded in ChPT will come into
play, and will regulate the behavior of the nuclear forces at
short distances. On the other hand, in quantum field theory, if we
spread interaction in space, we spread it in time as well. Thus
the processes that in describing low energy nucleon dynamics
should be used as a short-range part of the $NN$ interaction are
nonlocal in time. In the ordinary quantum mechanics nonlocality in
time of an interaction, when it is used as a "fundamental"
interaction generating the dynamics of a system creates the
conceptual problems with causality and unitarity. This is because
the Shr{\"o}dinger equation is local in time, and the interaction
Hamiltonian describes an instantaneous interaction. As we have
seen, the formalism of the GQD provides the extension of the
quantum dynamics needed for solving this problem, and the GDE
allows one to describe the evolution of a system with a
nonlocal-in-time interaction. It is remarkable that such a
nonlocalization can take place if and only if the UV behavior of
the matrix elements of the evolution operator is "bad".

As we have shown, the requirement that the $2N$ $T$-matrix
satisfies the GDE and has momentum behavior shown in Eq.
(\ref{anW}) allows one to construct it order by order  in the
effective expansion. Note that at any order $\langle {\bf
p}_2|T^{(2{\cal N})}(z)|{\bf p}_1\rangle$ constructed in this way
satisfies the GDE and hence is unitary. The corresponding
interaction operator describes the effective interaction that is
spread not only in space but also in time. This manifests itself
in the momentum and energy dependence of $\langle {\bf
p}_2|B^{(2{\cal N})}_{eff}(z)|{\bf p}_1\rangle$. The
regularization of the contact potentials spreads the interaction
described by this potential only in space. Iterating this
potential in the LS equation spreads the interaction also in time.
So in the energy region relevant for describing low energy nucleon
dynamics the effective interaction reproduced by the regularized
potential is nonlocal in time. However, since this potential gives
the main part of the contribution to the $T$-matrix only in the
region much above the scale $\Lambda_\chi$ where the
nonrelativistic two-nucleon LS equation is physically meaningless,
in this case the solution of the LS equation is only of formal
importance. To clarify this point note that the LS equation is
equivalent to the Schr{\"o}dinger equation that implies the
existence of the infinitesimal generator of the group of the
evolution operator $V(t)$. However, for the operator
$(V(t,0)-1)/t$ to approach to the limiting one sufficiently close,
one has to cross the region of times much below the time scale
$1/\Lambda_\chi$. For such times, as it follows from Eqs.
(\ref{U-G}) and (\ref{G-T}), the dominant contribution comes from
the operators   $T(z)$ corresponding to $z$ much above the scale
$\Lambda_\chi$. This means that the dynamics is not governed by
the Schr{\"o}dinger equation formulated in the terms of low energy
degrees of freedom. The above might be considered as an evidence
of inconsistency of the cutoff approach, but the formalism of GQD
provides a new reason to use the LS equation with a regularized
potential:The solution of the LS equation with any regularized
potential satisfies the condition (\ref{diferB2}), and hence the
solution for $z\in {\cal D}$ can be used as the effective
interaction operator in the GDE. In this case the existence of the
infinitesimal generator of the group of the evolution operators is
not required. However, as we have seen, not all the solutions of
the GDE relevant for the problem can be obtained in the above way.
In other words there are the interaction operators satisfying the
condition (\ref{diferB2}) that cannot be represented as a formal
solution of the LS equation with any regularized potential. The
above means that it is possible that not all the effects of
short-distance physics of QCD can be incorporated in the effective
theory of nuclear forces in the standard way where the short-range
part of the $NN$ interaction is described by the regularized
contact potentials. This motivates a modification of the chiral
potential model that could incorporate the description of the
short-range interaction based on the use of the GDE.

In the theory with pions the interaction operator should contain
terms describing both the short-range and long-range parts of the
$NN$ interaction. In Ref. \cite{FizB} it has been shown that at
leading order in the ${}^1S_0$ channel the $NN$ interaction
operator that is determined by the behavior of the $2N$ $T$-matrix
in the energy region ${\cal D}$ where the dominant contributions
to the $2N$ $T$-matrix come from the "$2N$ irreducible" diagrams
of ChPT  is of the form
\begin{eqnarray}\label{B^0_sh+}
\langle{\bf p}_2|B^{(0)}(z)|{\bf p}_1\rangle=\langle{\bf
p}_2|B^{(0)}_{sh}(z)|{\bf p}_1\rangle+\langle{\bf
p}_2|V^{(0)}_\pi|{\bf p}_1\rangle,
\end{eqnarray}
where the potential $V^{(0)}_\pi$ is given by Eq. (\ref{V_}),
i.e., is just the LO component of the long-range part of the
chiral potentials, and $B^{(0)}_{sh}(z)$ describes the short-range
component of the $NN$ interaction at LO and is given by Eq.
(\ref{B^0}). In the present paper we cannot discuss the theory
with pions in detail. However, the arguments presented in Ref.
\cite{FizB} lead us to the conclusion that at any order the $2N$
interaction operator can be represented as a sum of the
nonlocal-in-time short-range component being the sum of the
relevant contact diagrams and instantaneous part that is the sum
of the irreducible pion-exchange diagrams
\begin{eqnarray}\label{sh+pi}
\langle{\bf p}_2|B(z)|{\bf p}_1\rangle&=&\langle{\bf
p}_2|B_{sh}(z)|{\bf p}_1\rangle+\langle{\bf p}_2|V_\pi|{\bf
p}_1\rangle,
\end{eqnarray}
where the long-range component $\langle{\bf p}_2|V_\pi|{\bf
p}_1\rangle$ may be describe by that of the chiral potential at
the corresponding order.
 The solution of Eq. (\ref{difer}) with this interaction
operator is not more complicated than the solution of the LS
equation with an ordinary potential. In fact, the solution of Eq.
(\ref{difer}) with the boundary condition (\ref{T(z)to}) where
$B(z)$ is given by Eq. (\ref{sh+pi}) can be represented in the
form
\begin{eqnarray}\label{T_sT_pi}
T(z)=T_{sh}(z)&+&\Big(1+T_{sh}(z)G_0(z)\Big)\tilde
T_\pi(z)\nonumber\\
&\times&\Big(1+G_0(z)T_{sh}(z)\Big),
\end{eqnarray}
where $T_{sh}(z)$ is the $T$-matrix describing the dynamics by
$B_{sh}(z)$ alone, i.e., is the solution of equation
\begin{eqnarray}\label{difT_s}
\frac{dT_{sh}(z)}{dz}=-\int\frac{d^3k}{(2\pi)^3}\frac{T_{sh}(z)|{\bf
k}\rangle\langle {\bf k}|T_{sh}(z)}{(z-E_k+i0)^2},
\end{eqnarray}
with the boundary condition
$T_{sh}(z)\tend\limits_{|z|\tend\infty}B_{sh}(z)$, and $\tilde
T_\pi(z)$ satisfies the equation
\begin{eqnarray}\label{T_sT_pi}
\tilde T_\pi(z)=V_\pi+V_\pi G_{sh}(z)\tilde T_\pi(z),\nonumber\\
G_{sh}(z)=G_0(z)+ G_0(z)T_{sh}(z)G_0(z).
\end{eqnarray}
Note that, by using the "two-potential trick" \cite{Newton}, the
solution of the LS equation with the potential $V=V_{sh}+V_\pi$
can be written just in the same form. The only difference is that
in the latter case $T_{sh}(z)$ is the solution of the LS equation
with the potential $V_{sh}(z)$, while in the general case where
Eq. (\ref{T_sT_pi}) is derived from the GDE with the interaction
operator (\ref{sh+pi}), $T_{sh}(z)$ is the solution of Eq.
(\ref{T_sT_pi}) with the interaction operator $B_{sh}(z)$.

At $NNLO$ the $NN$ interaction in the ${}^1S_0$ channel can be
parametrized by the operator
\begin{eqnarray}\label{+pi}
\langle{\bf p}_2|B(z)|{\bf p}_1\rangle&=&\langle{\bf
p}_2|B^{(2)}_{eff}(z)|{\bf p}_1\rangle+\langle{\bf p}_2|V_\pi|{\bf
p}_1\rangle,\\
\langle{\bf p}_2|V_\pi|{\bf p}_1\rangle&=&\langle{\bf
p}_2|V^{(0)}_\pi|{\bf p}_1\rangle\nonumber\\&+&\langle{\bf
p}_2|V^{(2)}_\pi|{\bf p}_1\rangle+\langle{\bf
p}_2|V^{(3)}_\pi|{\bf p}_1\rangle,
\end{eqnarray}
where the operator $B_{eff}^{(2)}(z)$ is given by Eq. (\ref{B_at
NLO}), and $\langle{\bf p}_2|V^{(\nu)}_\pi|{\bf p}_1\rangle$ are
the pion-exchange components of the chiral potential
(\ref{chipot}) projected to the ${}^1S_0$ state. Thus for the
$NNLO$ $2N$ $T$-matrix in the ${}^1S_0$ channel we have
\begin{eqnarray}\label{T^1T_pi}
&&\langle{\bf p}_2|T(z)|{\bf p}_1\rangle=\langle{\bf
p}_2|\left(T_{sh}^{(2)}(z)+T_{sh}^{(2)}(z)G_0(z)\right)\tilde T_\pi(z)\nonumber\\
&\times&\left(1+G_0(z)T_{sh}^{(2)}(z)\right)|{\bf
p}_1\rangle\left(1+O\left(\left\{Q/\Lambda\right)^3\right\}\right),
\end{eqnarray}
where $T_{sh}^{(2)}(z)$ is given by Eq. (\ref{T2}). At the same
time, as we have shown, the result of
 resuming the contact potential (\ref{NLO}) to all orders can be
 represented in the form (\ref{T^2}) with the LEC's given by Eqs.
 (\ref{c_2}) and (\ref{J_1}). In other words, there are the solution sets
 at which one can arrive starting from the $LS$
 equation with the contact potential (\ref{NLO}), and in the case
 when $T_{sh}^{(2)}(z)$ in Eq. (\ref{T^1T_pi}) belongs to such sets
 this equation represents the $T$-matrix (more precisely a set of
 the $T$-matrices) that coincide with the $2N$ $T$-matrix obtained
 by solving the $LS$ equation with the $NNLO$ chiral potential.

 Thus  in the particular case the GDE with the interaction operator
 (\ref{+pi}) can reproduce the same results as the LS equation
 with the
 $NNLO$ chiral potential. However, Eq. (\ref{T^1T_pi}) that represents the solutions
 of the GDE with the interaction operator (\ref{+pi}) allows a
 much wider class of the solutions, and only experiment can
 discriminate between them.
  As we show in the next section, the above freedom in
 choosing values of
 the LEC's characterizing the short-range part of the $NN$
 interaction allows one to significantly change the off-shell
 predictions of the model and at the same time to keep the $NN$ phase shifts
 unchanged. This gives us the hope that the above modification of
 the chiral potential model could increase their predictive power in
 describing few-body and microscopic nuclear structure problems
 where the off-shell properties of the $NN$ interaction play an
 important role.

\section{Effects of the short-range forces on the off-shell \emph{2N T}-matrix and the \emph{pp} bremsstrahlung}

 One of the simplest processes involving the half
off-shell $NN$ interaction is proton-proton bremsstrahlung. Many
years ago, it was suggested to use the bremsstrahlung process as a
tool to discriminate between the various existing $2N$ potential
models \cite{Ashkin}. However, it has been shown that once all
ingredients which are necessary for the calculation of the
bremsstrahlung process are introduced, the predictions of various
modern potential models do not differ significantly, independent
of whether they are finite range or energy dependent
\cite{Herrmann}. The problem is that there is a significant
discrepancy between these predictions and experimental data of KVI
\cite{M-S} for asymptotic proton angles: As was shown in
Ref.\cite{M-S}, the microscopic calculations reproduce the
$pp\gamma$ cross sections reasonable well, however, for certain
kinematics for which the final $pp$ system has a low kinetic
energy, there is a significant over predictions of the data by
about $20\%-30\%$. Since the discrepancy between the results of
the microscopic calculations and the experimental results
increases as the kinetic energy of outgoing protons decreases, one
would speculate that including the Coulomb force might resolve the
problem. However, it has been shown by Cozma et al. \cite{C-T}
that this is not the case. An important part of this discrepancy
between theory and experiment originates in a poor description of
the $NN$ interaction in the ${}^1S_0$ channel at low energies
\cite{C-T}. However, the above refers only to potentials model,
while, as we have seen, there are such short-distance effects that
can be represented by the GDE but not by any potential model. This
leads us to the conclusion that the possible source of the
discrepancy between the data and the microscopic calculations is
the inability of the potential approach to incorporate such
effects. Let us now examine this possibility in the same way that
was used by Cozma et all \cite{C-T} for investigating the effect
of including the Coulomb interaction. In Ref. \cite{C-T} this
effect was studied by using the separable approximation to the
$NN$ interaction. The form factor $g({\bf p})$ in the separable
potential
\begin{equation}\label{potTjon}
  V({\bf p}_2,{\bf p}_1)=\lambda g^*({\bf p}_2)g({\bf
  p}_1)
\end{equation}
was chosen in the form
$$g({\bf p})=\frac{\beta^2}{p^2+\beta^2},$$
and the parameters $\lambda$ and $\beta$ were fixed by the
scattering length $a$ and the effective range $r_0$. The Coulomb
effect was investigated by adding the Coulomb potential to the
separable potential, and these corrections was shown to be minor
in the region needed for the KVI bremsstrahlung experiment.

Let us now, instead of the Coulomb effect, investigate the effects
of the including the short-range NLO interaction operator
$B^{(2)}_{sh}(z)$ given by Eq. (\ref{B_at NLO}) in combination
with the potential (\ref{potTjon}). In this case the interaction
operator of our problem is of the form
\begin{eqnarray}\label{Beff+Tjon}
\langle{\bf p}_2|B_{eff}(z)|{\bf p}_1\rangle=\langle{\bf
p}_2|B_{sh}^{(2)}(z)|{\bf p}_1\rangle\nonumber\\
+\lambda g^*({\bf p}_2)g({\bf
  p}_1).
\end{eqnarray}
For such a choice of the interaction operator, Eq. (\ref{T_sT_pi})
leads to the following $T$-matrix:
\begin{widetext}
\begin{eqnarray}\label{sol+Tjon}
\langle{\bf p}_2|T(z)|{\bf p}_1\rangle=\frac{\psi_0^*({\bf p}_2
)\psi_0({\bf
  p}_1 )t_{sh}(z)+g^*({\bf
p}_2)g({\bf
  p}_1)t_\pi(z)+\Big(\psi_0^*({\bf p}_2 )g({\bf
  p}_1)I^*_{\pi sh}(z^*)+g^*({\bf
  p}_2)\psi_0({\bf p}_1)I_{\pi sh}(z)\Big)t_{sh}(z)t_{\pi}(z)}{1-t_{sh}(z)t_{\pi}(z)I_{\pi sh}(z)I^*_{\pi sh}(z^*)}
\end{eqnarray}
with
\begin{eqnarray}
t_{sh}(z)&=&\frac{1}{C_0^{-1}-\frac{m}{4\pi}\sqrt{-zm}\Big(1+2\textsf{Re}c_2zm\Big)+{\cal
J}_2zm}\left(1+O\{Q^3/\Lambda^3\}\right),\nonumber\\
t_\pi(z)&=&\frac{1}{\lambda^{-1}-\int\frac{d^3k}{(2\pi)^3}\frac{|g({\bf
k})|^2}{z-E_k}},\nonumber\\
I_{\pi sh}(z)&=&\int\frac{d^3k}{(2\pi)^3}\frac{g({\bf
k})\psi^*_0({\bf
k})}{z-E_k}
=\Big[\frac{m}{4\pi}\frac{\sqrt{-zm}-\beta}{1+zm/\beta^2}\Big(1+c^*_2zm\Big)-{\cal
J}_{\pi sh}\Big]\Big(1+O\{Q^3/\Lambda^3\}\Big),\label{Jpi}
\end{eqnarray}
\end{widetext}
where
\begin{eqnarray}\label{J_pi}
{\cal J}_{\pi sh}=\frac{m}{2\pi^2}\int g({\bf
k})\Big(\psi^*_0({\bf k})-1\Big)dk.
\end{eqnarray}
 Here we restrict ourselves to the case
where the function $\psi_0({\bf p})$ that occur in Eq. (\ref{T2})
is real. Equation (\ref{sol+Tjon}) represents a set of the
solutions of the GDE that coincide with the true $2N$ $T$-matrix
with an accuracy of order $O\{Q^3/\Lambda^3\}$, because, as we
noted, the effective operator $B^{(2)}_{sh}(z)$ represents a set
of the operators describing the short-range part of the $NN$
interaction. Note that this solution set is determined not only by
the LEC's $C_0$, $c_2$, ${\cal J}_2$, $\beta$, and $\lambda$ that
occur in the interaction operator (\ref{Beff+Tjon}) but also by an
additional constant ${\cal J}_{\pi sh}$. This means that really
the operator (\ref{Beff+Tjon}) being the sum of the short-range
interaction operator and the potential $V_\pi$ is not so close to
the true solution in the domain ${\cal D}$ to uniquely determine
the set of the solutions of the GDE that coincides with an
accuracy of order $O\{Q^3/\Lambda^3\}$. In Ref. \cite{FizB} this
situation was explained by using the example of the interaction
operator (\ref{B^0_sh+}) generating low energy nucleon dynamics at
leading order. As it has been shown in Ref. \cite{FizB}, at high
$z$ this interaction operator is not so close to the true $2N$
$T$-matrix to uniquely determine the relevant solution. This means
that the operator (\ref{B^0_sh+}) does not satisfy the condition
(\ref{diferB2}). By substituting the operator (\ref{B^0_sh+}) into
Eq. (\ref{diferB2}), one can obtain terms that should be added to
the short-range part of the interaction operator \cite{FizB}. In
principle the interaction operator (\ref{Beff+Tjon}) should be
corrected in the same way. However, one may treat the interaction
operator of the form (\ref{Beff+Tjon}) keeping in mind that this
operator does not determine a unique solution of the GDE (more
precisely a unique solution set). Indeed, Eq. (\ref{sol+Tjon})
represents a family of the solution sets: In general the solutions
belonging to the sets with different values of the parameter
${\cal J}_{\pi sh}$ given by Eq. (\ref{J_pi}) do not coincide with
each other with an accuracy of order $O\{Q^3/\Lambda^3\}$. So to
select one of these sets we should chose the parameter ${\cal
J}_{\pi sh}$. In general this parameter should appear in the above
hidden additional terms, and the corrected interaction operator
should determine a unique solution (solution set) of the GDE. Thus
the set $\Omega_2$ of the solutions of the GDE that coincide with
the true $2N$ $T$-matrix with an accuracy of order
$O\{Q^3/\Lambda^3\}$ is determined by the LEC's $C_0$, $c_2$,
${\cal J}_2$,  ${\cal J}_{\pi sh}$ characterizing the short-range
part of the $NN$ interaction and the constant $\beta$, and
$\lambda$ parametrizing its long-range component. Each of the
solutions belonging to this set corresponds to some definite
function $\psi_0({\bf p} )$ satisfying the conditions
(\ref{Psi_0}), (\ref{J_2j}) and (\ref{J_pi}) with given LEC's
$c_2$, ${\cal J}_2$ and ${\cal J}_{\pi sh}$, and for momenta below
$\Lambda$ can be represented as
\begin{widetext}
\begin{eqnarray}\label{sol^2+Tjon}
\langle{\bf p}_2|T(z)|{\bf
p}_1\rangle&=&\Big[\frac{(1+c^*_2p_2^2)(1+c_2
p_1^2)t_{sh}(z)+g^*({\bf p}_2)g({\bf
  p}_1)t_\pi(z)}{1-t_{sh}(z)t_{\pi}(z)I^*_{\pi sh}(z^*)I_{\pi sh}(z))}\nonumber\\
 &+& \frac{\Big((1+c^*_2p_2^2)g({\bf
  p}_1)I^*_{\pi sh}(z^*)+g^*({\bf
  p}_2)(1+c_2
p_1^2)I_{\pi
sh}(z)\Big)t_{sh}(z)t_{\pi}(z)}{1-t_{sh}(z)t_{\pi}(z)I^*_{\pi
sh}(z^*)I_{\pi sh}(z)}\Big]\Big(1+O\{Q^3/\Lambda^3\}\Big).
\end{eqnarray}
\end{widetext}
The constants of the model  can be obtained from the requirement
that Eq. (\ref{sol^2+Tjon}) reproduces observables with an
accuracy of order $O\{Q^3/\Lambda^3\}$. Obviously the fitting to
the $pp$ scattering data is not sufficient for this, because these
constants determine not only the on-shell but also the off-shell
properties of the $2N$ $T$-matrix. For their determination the
comparison with experimental data such as the bremsstrahlung data
in which the off-shell properties of the $NN$ interaction manifest
themselves is needed.
 For example, one can use the $T$-matrix (\ref{sol^2+Tjon}) for
 describing the elastic $2N$ $T$-matrix in the $pp\gamma$ model of
 Martinus et al. \cite{{Martinus1},{Martinus2}}, and try to
 fit
the constants to reproduce the experimental cross sections.
However, in this paper we restrict ourselves to the investigation
of the effects of the short-range interaction on the off-shell
$2N$ $T$-matrix in the region needed for KVI bremsstrahlung
experiment.

In Figs. 4,5,6, and 7 the results of the calculations of the
$T$-matrix given by Eq. (\ref{sol^2+Tjon}).  The parameters
$\lambda$ and $\beta$ characterizing the long range component of
the interaction operator (\ref{Beff+Tjon}) were determined by
fitting the scattering amplitude reproduced by the pure potential
model to the scattering length $a=-18.1{\text{fm}}$ and the
effective range $r_0=2.60\text{fm}$. These values of $a$ and $r_0$
correspond to the case when the Coulomb interaction is switched
off \cite{C-T}.  The results of our calculations plotted in Figs.
4 and 5
 show that for some values of the LEC's  $c_2$, ${\cal J}_2$, and
${\cal J}_{\pi sh}$ the difference between the half off-shell
amplitudes $\langle{\bf p}_1|T(E_{p}+i0)|{\bf p}\rangle$ with and
without the short range corrections becomes larger as the momentum
going from the on-shell point, while, as we see from Fig. 4,
including the contact term keeps the phase shift unchanged. In
contrast, as has been shown in Ref. \cite{C-T}, the Coulomb
corrections become minor for high off-shell momenta, and hence
cannot remove the discrepancy between theory and experiment in
describing the $pp\gamma$ process. Our results show that with an
appropriate choice of the constants the short range part of the
$NN$ interaction gives a significant contribution to the $2N$
$T$-matrix in the case when the laboratory kinetic energy
$E_{\text{lab}}$ of the outgoing protons is below $15$ $MeV$ and
the off-shell momentum is sufficiently high. This is just the
region needed for the KVI bremsstrahlung experiment.
\begin{figure}
\resizebox{1\columnwidth}{!}{\includegraphics{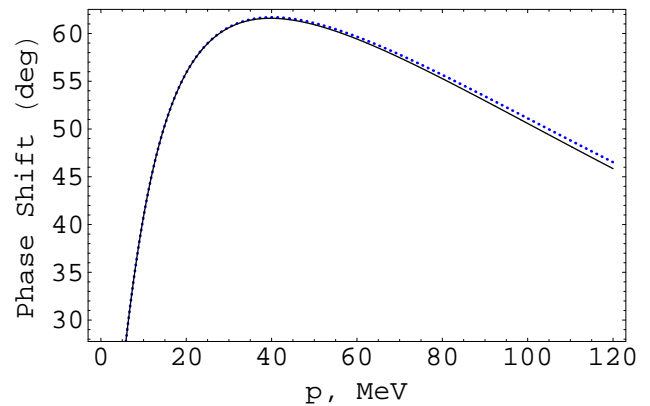}} \caption{A
comparison between the $pp$ phase shifts (${}^1S_0$ channel) for
the potential (\ref{potTjon}) fitted to $a=-18.1$ $\text{fm}$ and
$r_e=2.60$ $ \text{fm}$ (dotted line) and the predictions of the
model corrected by including the short effects (solid line).}
\end{figure}
\begin{figure}
\resizebox{1\columnwidth}{!}{\includegraphics{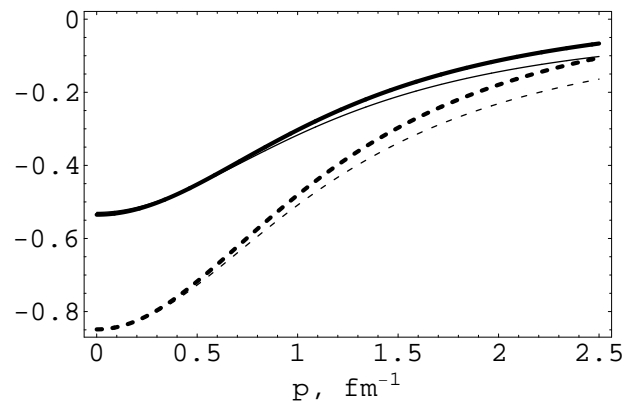}} \caption{The
$T$-matrix as a function of the off-shell momentum $p$ at a
laboratory energy of $10$ $\text{MeV}$. The solid and dashed
curves represent, respectively, the real and imaginary parts of
the normalized $T$-matrix $\langle
p|T_\pi(E_k+i0)|k\rangle/|\langle 0|T_\pi(E_k+i0)|k\rangle|$,
where $T_\pi(z)$ is the solution of the LS equation with the
potential (\ref{potTjon}). The heavy solid and bold type dashed
curves represent, respectively, the real and imaginary part of
$\langle p|T(E_k+i0)|k\rangle/|\langle 0|T_\pi(E_k+i0)|k\rangle|$,
where $T(z)$ is given by Eq. (\ref{sol+Tjon}).   The parameters
are following $\beta=241.468\text{MeV}$, $\lambda=-1.0269\cdot
10^{-4}\text{MeV}^{-2}$, $C_0=1.29\cdot 10^{-4}\text{MeV}^{-2}$,
${\cal
 J}_{\pi sh}=-17942{\text{MeV}}^{2}$, $c_2=2.3\cdot 10^{-7}\text{MeV}^{-2}$, ${\cal
 J}_2=2.75$.}
\end{figure}
\begin{figure}
\resizebox{1\columnwidth}{!}{\includegraphics{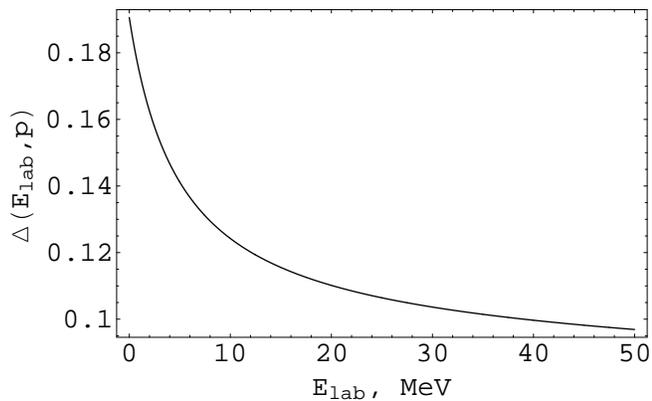}} \caption{The
difference $\Delta(E_{\text{lab}},p)$ between the predictions of
the model with  and without  the short-range corrections for the
T-matrix in the region needed for the KVI experiment ($p=300$
$\text{MeV}$).}
\end{figure}
\begin{figure}
\resizebox{1\columnwidth}{!}{\includegraphics{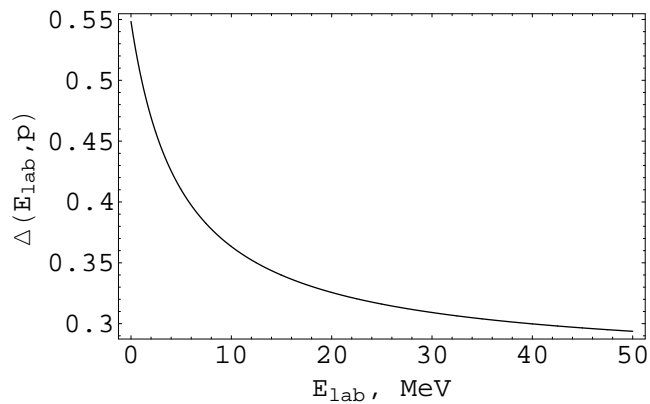}} \caption{Same
as Fig. 6 but for $p=500$ $\text{MeV}$.}
\end{figure}
A large discrepancy between experiment and the predictions of the
potential models appears in the kinematical regions where the
cross section has peaks. The presence of these peaks is the result
of strong final-state interaction at small values of the kinetic
energy of outgoing protons  ($E_{\text{lab}}<15\text{MeV}$). This
means that the over prediction of the potential models in
describing the $pp\gamma$ cross sections may be a reflection of
the fact that really the $NN$ interaction at low energies is not
so strong as it is predicted by the potential models. In Figs. 6
and 7 we plot the relative difference
\begin{eqnarray}
\Delta(E_{\text{lab}},p')=\frac{|\langle{\bf
p}'|T_\pi(E_p+i0)|{\bf p}\rangle|-|\langle{\bf p}'|T(E_p+i0)|{\bf
p}\rangle|}{|\langle{\bf p}'|T(E_p+i0)|{\bf p}\rangle|},\nonumber
\end{eqnarray}
$E_{\text{lab}}=2p^2/m$, between predictions of the model with and
without the short-range component for the half off-shell
$T$-matrix where $T_\pi(z)$ is the solution of the $LS$ equation
with the potential (\ref{potTjon}). As we see, for high off-shell
momenta, the magnitude of $\Delta(E_{\text{lab}},p)$
characterizing the effect of the short-range forces on the $2N$
$T$-matrix grows rapidly as the energy $E_{\text{lab}}$ going to
zero in the region $E_{\text{lab}}< 15$ $\text{MeV}$. Note that in
this kinematic region the difference between the predictions of
the microscopic models for bremsstrahlung and the KVI experimental
data has just the same energy dependence \cite{M-S}. This leads us
to the conclusion that the improvement of the low-energy part of
the strong interaction that can be achieved by correcting its
short-range component may remove the discrepancy between theory
and experiment. However, for including the short-range component
to give rise to the above effect, the constants $c_2$ and ${\cal
J}_2$ should have such values that are inconsistent with any
potential model. In fact, if the LEC's $c_2$ and ${\cal J}_2$
correspond to some contact potential then they must be related by
Eq. (\ref{J_1}). From this it follows that for $c_2=2.3\cdot
10^{-7}\text{MeV}^{-2}$ and ${\cal J}_2=2.75$ which lead to the
results shown in Figs. (4-7) the integral $\int d\tilde k|f(\tilde
k)|^2$ should be of order $\sim \frac{{\cal
J}_2\pi^2}{c_2m\Lambda_\chi}=1.26\cdot 10^5 \Lambda_\chi^{-1}$,
where the cutoff is set by the chiral symmetry breaking scale
$\Lambda_\chi=1GeV$. However, such a size of this integral is
inconsistent with the fact that the regulator function
$f(k/\Lambda)$ must satisfy $f(0)=1$ and fall off rapidly for
$k/\Lambda>1$. This means that the effective $2N$ $T$-matrix
(\ref{sol^2+Tjon}) corresponding to the above LEC's cannot be
associated with any potential.

The above model can be considered as a combination of the standard
nuclear physics approach based on the use of the $NN$ potentials
and the effective theory of nuclear forces. As we have seen,
combining our approach to the description of the short-range part
of the $NN$ interaction with the separable potential model yields
very encouraging results in investigating possible origins of the
discrepancy between the theoretical predictions for the $pp$
bremsstrahlung and experiment. In the same way one can construct a
more realistic model by replacing the potential (\ref{potTjon}) in
Eq. (\ref{Beff+Tjon}) with one of the high precision
phenomenological $NN$ potentials [25-30]. These potentials contain
the terms describing the short-range part of the $NN$ interaction,
and hence in this case the operator $B_{sh}(z)$ in Eq.
(\ref{B^0_sh+}) should be considered as a correction to the model
dependent short-range components of the realistic potentials in
which the ambiguous short-distance parametrization is used. It is
important that such corrections may allows one not only to remove
or minimize the model dependence but also improve the description
of the off-shell properties of the $NN$ interaction that is needed
for removing the discrepancy between the theory and the experiment
in describing the $pp$ bremsstrahlung.

\section{The off-shell behavior of the $2N$ $T$-matrix and
the three-nucleon problem}

 As we have shown, our formulation of the effective theory of nuclear forces
allows one to construct, as an inevitable consequence of the basic
principles of quantum mechanics and symmetries of QCD, not only
the $2N$ scattering amplitude (in this case we reproduce all
results of the standard subtractive EFT approach), but also the
off-shell $T$-matrix, and hence the evolution and Green operators.
The advantage of this formulation over the tradition potential
approach is that it allows one to take into account the
constraints on the off-shell properties of the $NN$ interaction
put by the symmetries of QCD. This is very important, because, as
is well known, the off-shell behavior of the $2N$ $T$-matrix may
play a crucial role in solving the many-nucleon problem and is an
important factor in calculating in-medium observables \cite{Fuchs}
and in microscopic nuclear structure calculations. This results,
for example, in the fact that the predictions by the Bonn
potential for nuclear structure problems differ in a
characteristic way from the ones obtained with local realistic
potentials \cite{Machleidt}.  The off-shell ambiguities of
realistic $NN$ potentials are argued to be one of the main causes
of many problems in describing three-nucleon systems.

The modern potential models  that describe $2N$ scattering data to
high precision can not guarantee that a similar precision will be
achieved in the description of larger nuclear systems. In fact,
the simplest observable in the $3N$ system, the binding energy of
the triton, is under predicted by the realistic $NN$ potentials
which are so successful in describing the $2N$ observables. The
energy deficit ranges from 0.5 to 0.9 MeV and depends on the
off-shell and short-range parametrization of the $2N$ force
\cite{Kievsky}. In order to resolve this problem one has to take
into account three-nucleon force (3NF) contributions to the $3N$
binding energy. The common way of solving the $3N$ bound state
problem is to use in the Schr{\"o}dinger equation phenomenological
$NN$ potentials and then to introduce a 3NF to provide
supplementary binding. However, from the point of view of the
three-nucleon problem, it is not sufficient to generate a
phenomenological $NN$ potential that perfectly reproduces the $2N$
scattering amplitudes. One must also generate a $NN$ potential by
using theoretical insight as much as possible in order to
constrain the off-shell properties of the $2N$ $T$-matrix. If this
is not the case, a $NN$ potential which fits precisely the $2N$
phase shifts but produces the erroneous off-shell behavior of the
$T$-matrix would not provide reliable results for the $3N$ system,
nor can be used to test for the presence of $3N$ forces.

The models for the $3N$F that are usually used for solving the
problem with the triton understanding are based on two-pion
exchange with intermediate $\Delta$-isobar excitation. However,
these $3N$F models cannot explain the $A_y$ puzzle which refers to
the inability to explain the nucleon vector analyzing power $A_y$
in elastic nucleon-deuteron ($Nd$) scattering at low energy. As
has been shown in Ref. \cite{Huber}, it is not possible with
reasonable changes in realistic $NN$ potentials to increase the
$Nd$ $A_y$ and at the same time to keep the $2N$ observables
unchanged. The same situation also takes place in the case of
chiral potentials \cite{Entem}. This means that the introduction
of three-nucleon forces (3NF) is needed for resolving the problem.
However, as it turned out, conventional $3N$F's change the
predictions for $Nd$ $A_y$ only slightly and do not improve them
[36-38]. This motivated new types of $3N$F's [39-41]. On the other
hand, reliable $3N$ calculations and even testing for the presence
of $3N$ forces require to constrain the off-shell properties of
the $2N$. Moreover, these properties play a crucial role in a new
$3N$F proposed by Canton and Schadow \cite{Canton}.
 In Ref.
\cite{Canton2} this $3N$F has been suggested as a possible
candidate to explain the $A_y$ puzzle. The $3N$F contribution from
this force is fixed by the $2N$ $T$-matrix describing the
underlying 2N interaction while the pion is "in flight". The
expression for this $3N$F  derived in Ref. \cite{Canton} contains
the off-shell $2N$ $T$-matrix, more precisely its subtracted part
\begin{equation}
  \tilde t_{12}({\bf p}_2,{\bf p}_1,z)=t_{12}({\bf p}_2,{\bf p}_1,z)-v_{12}({\bf p}_2,{\bf
  p}_1),\label{t_12}
\end{equation}
where ${\bf p}_1$ and ${\bf p}_2$ are Jacobi momenta of nucleons 1
and 2, and the potential - like term $v_{12}({\bf p}_2,{\bf p}_1)$
contains only OPE/OBE - type diagrams. The subtraction in Eq.
(\ref{t_12}) is needed to take into account a cancellation effect
which has been observed in Refs. \cite{Yang,Yang2}. This
cancellation involves meson retardation effects of the iterated
Born term, and the irreducible diagrams generated by sub-summing
all time ordering diagrams describing the combined exchange of two
mesons amongst the three nucleons. In principle there are no free
parameters to adjust, and the $3N$ force is completely determined
by the $2N$ $T$-matrix. However, as has been shown in Ref.
\cite{Canton2}, in order to explain the $A_y$ puzzle, instead of
the subtracted $T$-matrix shown in Eq. (\ref{t_12}), one has to
use the amplitude defined according to the prescription
\begin{eqnarray}
  \tilde t_{12}({\bf p}_2,{\bf p}_1,z)=c(z)t_{12}({\bf p}_2,{\bf p}_1,z)-v_{12}({\bf p}_2,{\bf
  p}_1),\nonumber
\end{eqnarray}
with the effective parameter $c(z)$, which represents an overall
correction factor for the far-off-shell $2N$ $T$-matrix. Ideally,
this parameter should be one for the $2N$ potential to provide a
reliable extrapolation of the $2N$ $T$-matrix down to $z\approx
-160$ MeV. However, as has been shown in Ref. \cite{Canton2}, none
of the existing $2N$ $T$-matrices can guarantee the off-shell
behavior that is needed for the explanation of the $A_y$ puzzle
with the parameter $c(z)$ set to one, since they are all
constrained by such data the deuteron pole and the $2N$ phase
shifts. For example, in order to reproduce the $nd$ experimental
data with the Bonn B potential, the factor $c(z)$ must be set to
$0.73$ for the energy $10$ $MeV$.

The above gives us the expect that the origin of discrepancy
between the theory and experiment in describing $A_y$ is the same
as in the case of the $pp$ bremsstrahlung: The potential models
cannot reproduce the relevant off-shell behavior of the $2N$
$T$-matrix. This should mean that the answer lies beyond these
models. The advantage of the formulation of the effective theory
is that it allows one to constrain the possible $T$-matrix by the
symmetries of QCD, but at the same time, permits the solutions
that cannot be reproduced by any potential model. This opens new
possibilities to resolve the $A_y$ puzzle. Note that the
generalization of the above results to the $P$ waves needed for
solving this problem is straightforward.

\section{Summary and Discussion}

We have shown that from the Weinberg analysis of time-ordered
diagrams for the $2N$ $T$-matrix in ChPT it follows that nucleon
dynamics at low energies is governed by the GDE with a
nonlocal-in-time interaction operator. A remarkable feature of the
GDE which follows straightforwardly from the first principles of
quantum mechanics is that it allows one to construct all physical
amplitudes relevant for the theory under consideration by using
the amplitudes describing processes in which the duration time of
interaction is infinitesimal. It is natural to assume that the
most of contribution to these amplitudes comes from the processes
associated with a fundamental interaction in a quantum system.
This point manifests itself in the boundary condition
(\ref{fund}). If we do not consider a theory that is valid up to
infinitely high energies (infinitesimal times), then the duration
times of interaction  much below the time scale of low energy
physics but  much above the scale of underlying high energy
physics should be considered as "infinitesimal", and the operator
describing the processes with such duration times of interaction
can be used as an interaction operator.
 This in turn means that the
amplitudes describing this "fundamental" interaction in the low
energy theory can be computed in terms of the underlying high
energy physics.

The GDE can be represented in the form of the differential
equation (\ref{difer}) for the operator $T(z)$. The boundary
condition on this equation is shown in Eq. (\ref{T(z)to}), where
the operator $B(z)$ describes the fundamental interaction in this
system. By definition, this operator must be so close to the true
$T$-matrix in the limit $|z|\to\infty$ that the GDE with the
boundary condition (\ref{T(z)to}) have a unique solution. The
above means that really this region of "infinite" energies with
which we have to deal is the domain ${\cal D}$ that lies much
above the scale of the low energy dynamics but much below the
scale of the underlying high energy physics. Correspondingly for
describing the low energy dynamics we have to start with the
boundary condition (\ref{T(z=s)}) that implies that the
interaction operator $B(z)$ is so close to the relevant $T$-matrix
inside the domain ${\cal D}$ that Eq. (\ref{difer}) with this
initial condition has a unique solution. In the theory of nuclear
forces this domain is a region of energies that are high enough
for the most of contribution to the $2N$ $T$-matrix to come from
processes that are described by the irreducible $2N$ diagrams for
the $2N$ $T$-matrix (these processes are associated with a
"fundamental" interaction), but not so high for the heavy degrees
of freedom manifest themselves explicitly. By using the analysis
of the time-ordered diagrams for the $2N$ $T$-matrix in ChPT
inside the domain ${\cal D}$ where the structure of the theory is
much simpler than in the low energy region, we can obtain the $NN$
interaction operator which should be used in the boundary
condition (\ref{T(z=s)}) on Eq. (\ref{difer}).  We have shown that
the dynamics which is generated by the interaction operator
obtained in this way is non-Hamiltonian. This is because, for low
energy nucleon dynamics to be Hamiltonian, the operator $T(z)$
must have a negligible dependence on $z$ inside the domain ${\cal
D}$, while, as we have shown, this is not the case. In other
words, in the nonrelativistic limit QCD leads through ChPT to low
energy nucleon dynamics that is not governed by the
Schr{\"o}dinger equation. However, this does not mean that the low
energy predictions of QCD are not consistent with quantum
mechanics: Only the GDE must be satisfied in all the cases. We
have shown that the requirement that the $2N$ $T$-matrix satisfies
the GDE as a well defined equation free from UV divergences and
has the structure consistent with the symmetries of QCD allows one
to construct it order by order in the effective theory expansion
without resorting to regularization and renormalization.

The above point is illustrated in detail by using the pionless
theory of nuclear forces. For the ${}^1S_0$ channel $2N$
$T$-matrix that, as it follows from the Weinberg analysis, should
be of the form (\ref{anW}), we have derived Eq. (\ref{dif_0up}).
This equation allows one to obtain the $2N$ $T$-matrix with an
accuracy of order $O\{(Q/\Lambda)^{2{\cal N}+1}\}$ if it is known
with an accuracy of order $O\{(Q/\Lambda)^{2{\cal N}-1}\}$. In
this way at any order one can obtain the most general possible
$2N$ $T$-matrix consistent with the symmetries of QCD. At NLO this
$T$-matrix is given by Eq. (\ref{T2}). This equation represents
the set $\Omega_2$ of the solutions of the GDE that coincide with
the true $2N$ $T$-matrix with the accuracy up to order
$O\{Q^3/\Lambda^3\}$. This solution set is generated by the
effective interaction operator (\ref{B_at NLO}). The fact that we
have to deal with the set of the solutions means that at this
order our knowledge about the details of the $NN$ interaction is
not sufficient for discriminating between them. The set $\Omega_2$
is determined by the LEC's $C_0$, $c_2$, and ${\cal J}_2$ that
parametrize the effects of the high energy physics on low energy
nucleon dynamics. It is shown that, the $2N$ $T$-matrix obtained
in this way leads to the KSW expansion of the scattering amplitude
\cite{Kaplan2} shown in Eq. (\ref{A^4}). However, in contrast with
the standard subtractive approach to the EFT of nuclear forces
where this expansion is obtained by summing the bubble diagrams
and performing regularization and renormalization, we reproduce
the same results in a consistent way free from UV divergences. At
NNLO the $2N$ $T$-matrix is given by Eq. (\ref{T^4()}). This
equation represents the set $\Omega_4$ of the solutions of the GDE
that coincide with the true $2N$ $T$-matrix with an accuracy of
order $O\{Q^5/\Lambda^5\}$. In the same way one can obtain the
solution set $\Omega_{2{\cal N}}$ that coincide with the true $2N$
$T$-matrix with the accuracy up to order $O\{(Q/\Lambda)^{2{\cal
N}+1}\}$. As ${\cal N}$ increases the set $\Omega_{2{\cal N}}$
becomes smaller and smaller, i.e., one approaches to the true $2N$
$T$-matrix closer and closer.

The above strategy can be also used for constructing the
$T$-matrix in the theory with pions. Another way of applying the
results obtained in the paper is to use them in combination with
the chiral potential approach. We have shown that at NNLO the $NN$
interaction in the ${}^1S_0$ channel can be parameterized by the
operator (\ref{+pi}). The solution of the GDE with this
interaction operator is given by Eq. (\ref{T^1T_pi}). In the
particular case when the short-range part of the $2N$ $T$-matrix
$T_{sh}^{(2)}(z)$ that enters in Eq. (\ref{T^1T_pi}) can be
associated with some potential model this equation leads to the
same results as the chiral model with the NNLO potential. At the
same time, as we have seen, there are such LEC's characterizing
the short-range part of the interaction operator (\ref{+pi}) in
the case of which $T_{sh}^{(2)}(z)$ cannot be associated with any
potential model, and Eq. (\ref{T^1T_pi}) gives rise to the $2N$
$T$-matrix that cannot be reproduced by the LS equation with the
NNLO chiral potential.
 Thus, being equivalent to the chiral potential model in a
particular case, above model allows one to go beyond the potential
models, and such an extension of the standard chiral approach may
be necessary to remove the discrepancy between the theory and
experiment still existing in nuclear physics. This point has been
illustrated by using the example of the model where the long-range
part of the interaction is described by the potential
(\ref{potTjon}). This is the model that in Ref. \cite{C-T} was
used for investigating the effects of including the Coulomb
interaction on the $pp\gamma$ process where the Coulomb potential
is replaced by the short-range interaction operator
(\ref{Beff+Tjon}). By using this model we have shown that, as we
see from Figs. 6 and 7, there are such LEC's for which the effect
of the short-range interaction on the half off-shell $T$-matrix
$\langle{\bf p}_1|T(E_p+i0)|{\bf p}\rangle$ grows rapidly as the
kinetic energy of outgoing protons decreases  in the region needed
for KVI experiment. Note that just for such LEC's the short-range
interaction that is described by Eq. (\ref{T2}) cannot be
associated with any potential. Such an effect of including the
short-range interaction might seem very surprising, but the
difference between the predictions of the potential models and
experiment in this kinematic region grows in the same way. This
lead us to the conclusion that the inability of the potential
models to incorporate the short-distance effects which manifest
themselves when we consider the problem from the point of view of
the more general approach based on the use of the GDE, may be the
source of the over prediction of the potential models in
describing the $pp\gamma$ cross section.  As we see from Figs.
4,5,6, and 7, these effects can improve the off-shell behavior of
the $2N$ $T$-matrix and the same time keep the phase shift
unchanged. This allow one to incorporate the above effects into
the machinery based on the use of the high precision potentials
keeping the phase shifts predicted by these potentials unchanged.
As we noted, in this case the operator $B_{sh}(z)$ should describe
corrections to these short-range components. One may hope that
such an improvement of the description of the $NN$ interaction
will make it possible to remove the discrepancy between theory and
experiment in describing the $pp$ bremsstrahlung. Taking into
account the above effects of the short-range interaction on the
off-shell properties of the $2N$ $T$-matrix may also open new
possibilities for explaining the $A_y$ puzzle.

\end{document}